\documentclass[a4paper,11pt]{article}
\pdfoutput=1
\usepackage{jcappub}
\usepackage[T1]{fontenc}
\usepackage[utf8]{inputenc}
\usepackage{mathrsfs}
\usepackage{bm}
\usepackage{url}
\usepackage[normalem]{ulem}
\usepackage{mathtools}
\usepackage{array}
\newcolumntype{P}[1]{>{\centering\arraybackslash}p{#1}}
\newcolumntype{M}[1]{>{\centering\arraybackslash}m{#1}}
\usepackage[caption=false]{subfig}
\usepackage{dcolumn}
\usepackage{graphicx, epsfig}
\usepackage[dvipsnames]{xcolor}
\usepackage{mathrsfs}
\usepackage{bm}
\usepackage{yhmath}
\usepackage[caption=false]{subfig}
\usepackage[normalem]{ulem}
\usepackage{mathtools}
\usepackage{bigints}
\usepackage{float}
\usepackage{hyperref}

\def\t13{\mathrel{{\theta_{13}}}}
\def\y12{\mathrel{{\tan^2 \theta_{12}}}}
\def\c2{\mathrel{{\chi^2 }}}

\newcommand{\be}{\begin{equation}}
\newcommand{\ee}{\end{equation}}
\newcommand{\ba}{\begin{eqnarray}}
\newcommand{\ea}{\end{eqnarray}}

\title{
Observational strategies for ultrahigh-energy neutrinos: \\the importance of deep sensitivity for detection and astronomy
}
\author[a,b,c,1]{Kumiko Kotera,\note{Corresponding author.}}
\author[c]{Mainak Mukhopadhyay,}
\author[d,a]{Rafael Alves Batista,}
\author[c]{Derek Fox,}
\author[e]{Olivier Martineau-Huynh,}
\author[c,f]{Kohta Murase,}
\author[c]{Stephanie Wissel,}
\author[c]{and Andrew Zeolla}
\affiliation[a]{Sorbonne Universit\'e, CNRS, UMR 7095, Institut d’Astrophysique de Paris, 98 bis bd Arago, 75014 Paris, France
}
\affiliation[b]{Astrophysical Institute, Vrije Universiteit Brussels, Pleinlaan 2, 1050 Brussels, Belgium
}
\affiliation[c]{Department of Physics; Department of Astronomy \& Astrophysics; Center for Multimessenger Astrophysics, Institute for Gravitation and the Cosmos, The Pennsylvania State University, University Park, PA 16802, USA.}
\affiliation[d]{Universite Paris Diderot, Sorbonne Paris Cite, CNRS Laboratoire de Physique Nucleaire et de Hautes Energies, 4 place Jussieu F-75252, Paris Cedex 5, France
}
\affiliation[e]{Sorbonne Universit\'e, CNRS, Laboratoire de Physique Nucléaire et des Hautes Energies (LPNHE), 4 Pl. Jussieu, 75005 Paris, France
}
\affiliation[f]{Center for Gravitational Physics and Quantum Information, Yukawa Institute for Theoretical Physics, Kyoto, Kyoto 606-8502 Japan}
\emailAdd{kotera@iap.fr}
\emailAdd{mkm7190@psu.edu}
\abstract{
Detecting ultrahigh-energy neutrinos can take two complementary approaches with different trade-offs.  1)~Wide and shallow: aim for the largest effective volume, and to be cost-effective, go for wide field-of-view but at the cost of a shallow instantaneous sensitivity -- this is less complex conceptually, and has strong discovery potential for serendipitous events. However, it is unclear if any source can be identified, following detection. And 2)~Deep and narrow: here one uses astrophysical and multi-messenger information to target the most likely sources and populations that could emit neutrinos -- these instruments have deep instantaneous sensitivity albeit a narrow field of view. Such an astrophysically-motivated approach provides higher chances for detection of known/observed source classes, and ensures multi-messenger astronomy. However, it has less potential for serendipitous discoveries. In light of the recent progress in multi-messenger and time-domain astronomy, we assess the power of the deep and narrow instruments, and contrast the strengths and complementarities of the two detection strategies. We update the science goals and associated instrumental performances that envisioned projects can include in their design in order to optimize discovery potential.
}
\begin{document}
\maketitle
\flushbottom
\section{\label{sec:level1}Introduction}
Ultrahigh-energy (UHE, energy $> 10^{17}\,$eV) neutrinos are daughter particles of the routinely detected ultrahigh energy cosmic rays (UHECRs) that interact with cosmic photons or baryons. These interactions can happen in the source environment of UHECRs, while they are accelerated or are escaping -- these are called {\it astrophysical} neutrinos. Alternatively, they can take place in the intergalactic medium, during the flight of these particles from their sources to the Earth -- these are called {\it cosmogenic} neutrinos. Because these sources have been confirmed to be of extragalactic origin \cite{2017Sci...357.1266P} and because of our increasingly precise knowledge of the extragalactic background light, the existence of UHE neutrinos at some flux level is guaranteed. 

UHE neutrinos have been searched for in the past three decades, first as an ancillary science case of ultrahigh energy cosmic ray (UHECR) instruments (e.g., the Pierre Auger Observatory, \cite{Aab_2019}) and high-energy (PeV) neutrino detectors (e.g., IceCube,~\cite{IceCubeCollaborationSS:2025jbi}). These instruments have derived upper limits on the fluxes of UHE neutrinos, ruling out exotic UHECR production models and the most optimistic UHECR source scenarios. 
Dedicated instruments at ultrahigh energies have also been built, with balloons (ANITA), and more recently, in-ice radio detectors at the South Pole or in Greenland (ARA, ARIANNA, RNO-G) -- see \cite{decoene2023reviewneutrinoexperimentssearching} for a review.

\begin{figure*}[!t]
\centering
\includegraphics[width=0.8\linewidth]{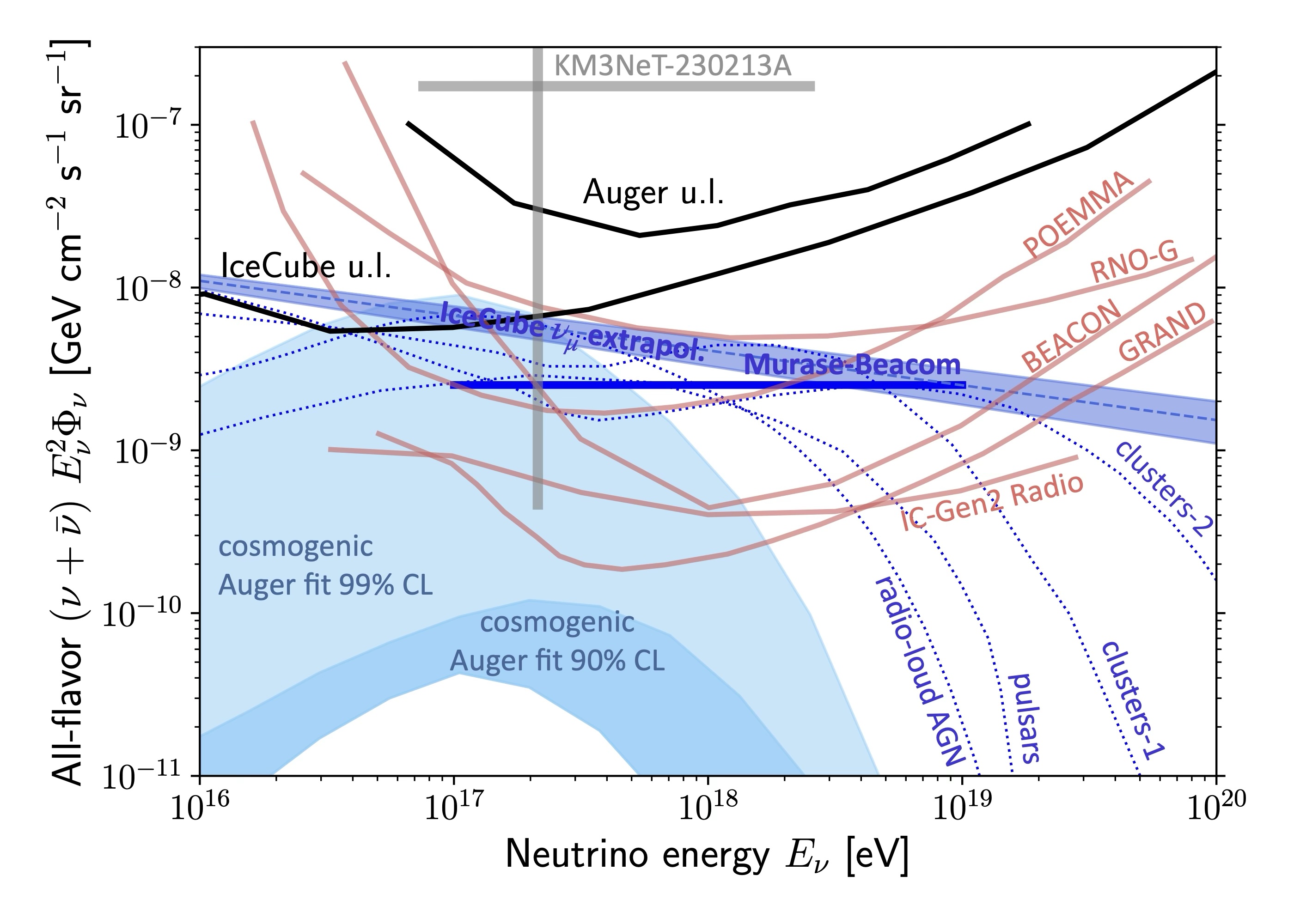}
\caption{What diffuse flux level should we aim for at ultrahigh energies? Theoretically predicted diffuse UHE neutrino fluxes from astrophysical (blue dotted lines, see text) \cite{Kotera09, PhysRevD.90.103005, Fang:2017zjf, Rodrigues21, lundquist2024} and cosmogenic origins, obtained with a comprehensive fit to the Auger UHECR data: 99\% C.L. fit (light+darker blue band), corresponding to the standard source parameters and 90\% C.L. fit (darker blue band), for pessimistic parameters~\cite{Alves_2019}. Overlayed is a theoretical extension to UHE energies of the measured IceCube flux~\cite{Stettner:2019tok,IceCube21} (navy blue band) and the Murase-Beacom (or effective nucleus-survival) line (thick navy blue line): the astrophysical neutrino flux level expected for iron UHECRs, assuming an effective photodisintegration optical depth of 1, and strong source evolution history \cite{Murase_2010}.
The pink solid lines indicate the projected 10-year differential sensitivities of several projects (for BEACON, the 1000 station limit is quoted~\cite{Wissel:2020sec, BEACON:2025qcq}). Black solid lines mark the  upper limits on UHE neutrinos from IceCube \cite{IceCubeCollaborationSS:2025jbi}
and Auger \cite{Aab_2019}. The gray cross corresponds to the flux needed to achieve the detection of the event KM3-230213A. The horizontal span corresponds to the central 90\% neutrino energy range associated with the event, and the vertical bars represents the $3\sigma$ Feldman–Cousins confidence interval on this estimate \cite{KM3NeT:2025npi}. Note that the IceCube limits are derived assuming an $E^{-1}$ differential flux, whereas the Auger limits assume an $E^{-2}$ flux, and are therefore not directly comparable.}

\label{fig:diffuse}
\end{figure*}

The recent progress in relation to this search can be summarized as follows:

\begin{enumerate}
\item The Auger observations of UHECRs constrain the flux of cosmogenic neutrinos towards low fluxes \cite{Alves_2019}.

\item The advent of multi-messenger astronomy and progress in time domain astronomy has revolutionized high-energy astrophysics. Notions of target of opportunity, alert systems, pointing, etc. require specific strategies of observations and hence novel instrumental performances \cite{2022NatRP...4..697G}.

\item The IceCube Observatory has detected astrophysical neutrinos at TeV-PeV energies \cite{2013PhRvL.111b1103A,2013Sci...342E...1I,2020PhRvL.124e1103A}. Two sources have been clearly identified after 10 years of neutrino event accumulation \cite{IC_Science_2022, IceCube_Science_2023} and there are other hints. 

\item One UHE neutrino has likely been detected by KM3NeT, but its origin is unclear~\cite{KM3NeT:2025npi} (see more details below.)
\end{enumerate}

Upon its first phase of data taking, KM3NeT detected an isolated remarkable near-horizontal muon, with with an estimated energy of $\sim 120^{+110}_{-60}\,$PeV. In spite of systematic and statistical uncertainties, the probability that the muon could have originated from an atmospheric cosmic ray shower is low. However, the volume of the deployed instrument at the time of detection challenges a diffuse flux origin of this event, the 10 times larger IceCube not having detected any such neutrinos. In case of a serendipitous transient event, the absence of any powerful source nor obvious host galaxy in the particle arrival direction further challenges the interpretation of this detection \cite{KM3NeT:2025npi,KM3NeT:2025aps,KM3NeT:2025bxl,KM3NeT:2025ccp,KM3NeT:2025vut}. This puzzling detection highlights the necessity of clarifying the astrophysical expectations and the detection strategies in this extreme regime.

Today, the construction of very large scale instruments is envisioned, with on-going prototyping phases for detection technique validation (e.g., BEACON~\cite{Wissel:2020sec,Southall:2022yil}, GRAND~\cite{GRAND:2018iaj}, IceCube-Gen2 Radio~\cite{2020arXiv200804323T}, POEMMA or SPB-2~\cite{2021cosp...43E1367O}, PUEO~\cite{2021JInst..16P8035A},  TAMBO~\cite{2020arXiv200206475R, TAMBO:2025jio},  Trinity~\cite{2021arXiv210802751W}). 
We have a window of a few years to orient the design and conception of these instruments before full deployment, and optimize complementary strategies. 

The key question that we seek to address in this work is: if one or several instruments could be funded from several sources (as was the case for IceCube and ANTARES or KM3NeT), what would be a winning strategy, in terms of site distribution on the Earth, and in terms of techniques? And directly linked to this question, what will then be the field of view (FoV), sensitivity (instantaneous, daily averaged, diffuse), angular resolution that could be globally achieved?\\

All envisioned experiments are planning to reach excellent sensitivity to the diffuse flux over large portions of the sky (typically a 10-year integrated diffuse sensitivity limit $\sim 10^{-10}\,{\rm GeV}\,{\rm cm}^{-2}\,{\rm s}^{-1}\,{\rm sr}^{-1}$ above $5\times 10^{17}$\,eV), and their performances align on this number (see Table~1 of \cite{2022NatRP...4..697G}). 

However, these detectors are intrinsically different, and can be classified in two categories: 
\begin{enumerate}
\item instruments with wide FoV and shallower sensitivity
\item instruments with deeper sensitivities and narrow FoV.
\end{enumerate}

One can draw an obvious parallel with more classical astronomical observatories. Both types of instruments are needed for efficient astronomical observations, but their use and science case have to be tuned to their capabilities.\\ 

\begin{figure}
\centering
\includegraphics[width=0.8\columnwidth]{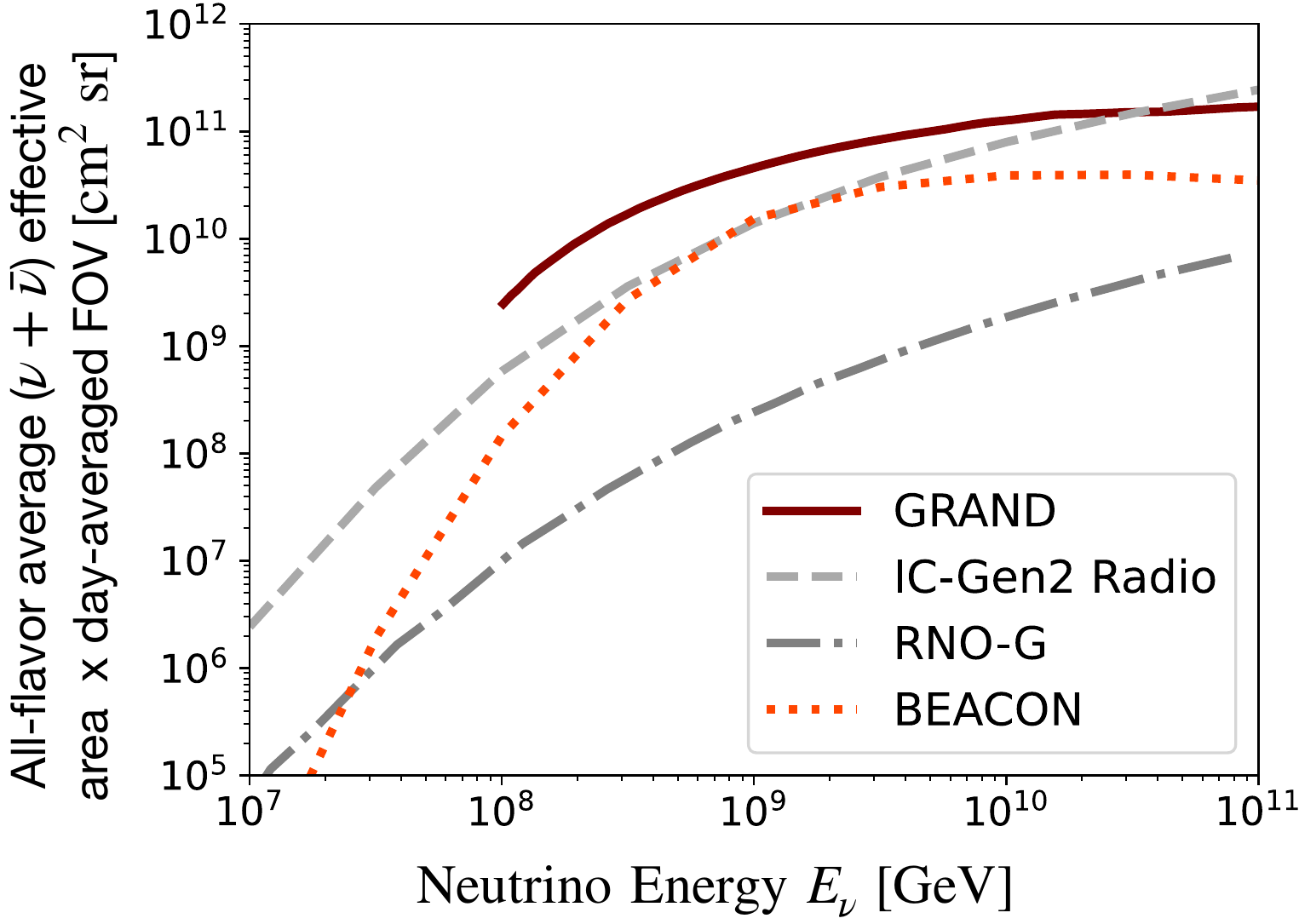}
\includegraphics[width=0.8\columnwidth]{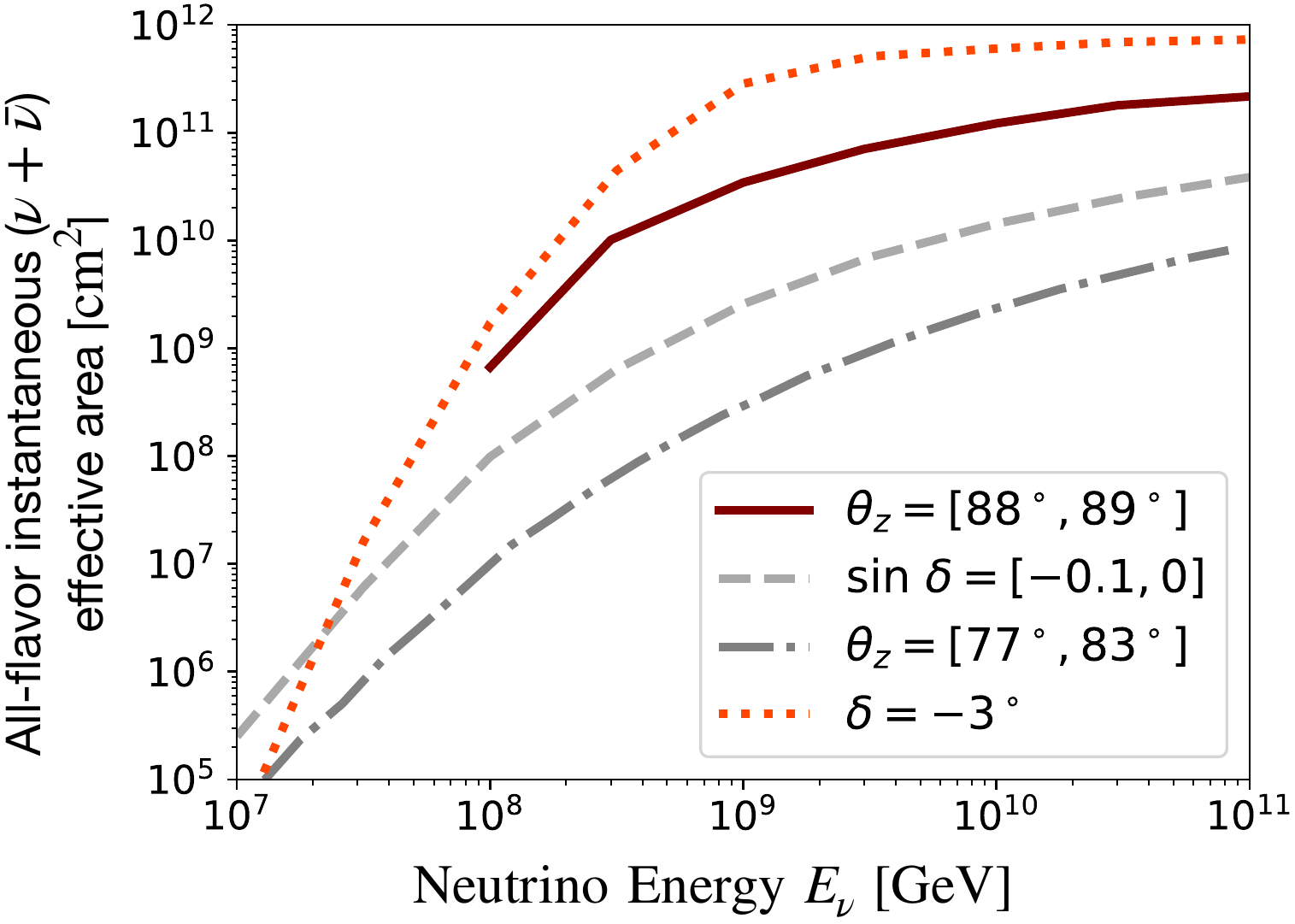}
\caption{\emph{Top: }Acceptance, defined as the product of the average effective area and the day-averaged field-of-view (FOV) in steradians, shown for GRAND~\cite{GRAND:2018iaj} (solid), IceCube-Gen2 Radio~\cite{IceCube-Gen2:2021rkf,Gen2_TDR,Glaser:2019cws} (dashed), and RNO-G~\cite{RNO-G:2020rmc} (dot-dashed), and BEACON~\cite{BEACON:2025qcq} (dotted). \emph{Bottom: }The instantaneous effective area for each detector in the most sensitive band, where $\theta_z$ denotes zenith angle and $\delta$ denotes declination. Note that the legend is the same as above with the most sensitive band indicated for each detector.
}
\label{fig:avg_effectiveArea}
\end{figure}

In this paper, we examine the major science cases and detection strategies that each of these types of instruments can target, in view of our current knowledge of the high-energy (HE) astrophysical source populations. By reviewing various points of observation strategy, we aim at casting light on the pros and cons of various detection techniques and associated performances, so as to inform the design choices of up-coming experiments. We focus in particular on the power of the deep and narrow instruments in detecting UHE neutrinos and performing astronomy.

In Section~\ref{section:diffuse}, we present an update on the diffuse UHE neutrino fluxes that are most realistic and relevant to aim for. In Section~\ref{section:transient}, we make a census of the rates and energies of the transient populations capable of accelerating UHE neutrinos and their associated detection potential with different types of instruments, according to their duration. Finally, in Section~\ref{section:nearby}, we examine the most likely host galaxies for UHE neutrino transients in the Local Universe. We conclude and outline our perspectives in Section~\ref{sec:concl}.

\section{Diffuse fluxes and identifying point sources within}\label{section:diffuse}

Up to TeV-PeV energies, the IceCube Observatory has detected a diffuse astrophysical neutrino flux. Its spectrum is best fit with a single power-law in $E_\nu^{-\gamma}$, where $\gamma=2.37$ for the 9.5 years of muon neutrinos \cite{IceCube:2021uhz} and $\gamma=2.87$ for 7.5 years of HESE events \cite{IceCube21} which consist of both tracks and cascades. Considering most of the current models, a cosmogenic origin for this flux is likely ruled out at these low energies and high flux levels. The astrophysical origin of the bulk of the flux is not identified yet. Two sources have been identified: an event excesses associated with NGC\,1068 with a significance of $4.2\,\sigma$ \cite{IC_Science_2022}, and an emission from the Galactic plane at $4.5\,\sigma$ level of significance, consistent with a diffuse emission from the Galactic plane or a population of unresolved point sources \cite{IceCube_Science_2023}. Furthermore, there is a $3\,\sigma$ evidence of a spatially coincident neutrino event with the blazar TXS 0506+056~\cite{IceCube:2018dnn,IceCube:2018cha} and hints of possible neutrino correlations with several tidal disruption events~\cite{Stein20,Reusch:2021ztx,vanVelzen:2021zsm,Mukhopadhyay:2023mld}.

The level of the observed diffuse flux is similar to the so-called nucleon-survival line derived by Waxman and Bahcall, a neutrino flux level derived under the assumption that they are secondary particles of detected UHECRs (assuming a light composition), produced in optically thin extragalactic sources \cite{WaxmanBahcall98}. For decades, that flux had been used as a  benchmark level to aim for experimentally. 

At ultrahigh energies, so far, the prominent strategy has been to look for cosmogenic UHE neutrinos, as a guaranteed secondary flux to observed UHECRs. An extension of the nucleon-survival line to UHE also used to be a target, but it is already ruled out by the IceCube and Auger non-detections \cite{Aartsen_2018,Aab_2019,gonzalez2022,AbdulHalim:2023SN}. Because of the four recent developments highlighted in the introduction, it is time to revisit the goals of any search for UHE neutrinos. 

\subsection{Going beyond the cosmogenic flux: what is a new diffuse flux level to aim for?}

\subsubsection{Updated cosmogenic flux levels} 
A first remark is that the UHE cosmogenic neutrino flux is severely constrained by the UHECR observations by the Auger Observatory, to below 
\begin{equation}
E_\nu^2\Phi_{\rm cosmo, max} \,\sim 10^{-8}\,{\rm GeV}\,{\rm cm}^{-2}\,{\rm s}^{-1}\,{\rm sr}^{-1}\,{\mbox{at } E_\nu \simeq}\ 3\times 10^{17}\,{\rm eV}\ , \,\mbox{ at } \, 99\% \mbox{ C.L.}\ .
\end{equation}

The combined fit of the spectrum and composition data of the Auger Observatory \cite{Aab_2017,PierreAuger:2022atd,PierreAuger:2023htc} points towards non-light composition and hard injection spectra at the sources, that do not favor abundant cosmogenic neutrino flux production. 

An updated combined fit, relaxing the source population evolution history as a free parameter, was performed, and the corresponding cosmogenic UHE neutrino fluxes was derived in \cite{Alves_2019}. This study showed that, at 99\% C.L., the flux should be contained in the blue band (light+darker) presented in Figure~\ref{fig:diffuse}. These contain all standard models of UHECR sources, with their conservative parameters, that allow for a 99\% CL. fit to the Auger data. More aggressive scenarios with hard spectra and negative source evolutions yield the 90\% C.L. darker blue band at the bottom of the cosmogenic blue bland. An independent analysis performing a fit to Auger data was conducted by \cite{Heinze_2019} and lead to comparable results. A more recent fit to the Auger data points towards fluxes in the lower range of the blue band, grazing the best instrumental sensitivity lines \cite{PierreAuger:2022atd}.

These calculations however have a limitation in the fact that there is a degeneracy between the source number density and luminosity evolutions, such that the redshift dependence should take into account both of these quantities, following the Peters cycle (see also \cite{muzio2024peterscycleendcosmic}). Although the redshift dependency is treated separately for both quantities in these studies, \cite{Ehlert_2023} demonstrates that accounting properly for the redshift evolution would degrade the fit to UHECR data, and/or lead to lower cosmogenic neutrino flux levels. On the other hand, it is also viable that a small fraction of protons in the UHECR flux could still generate a significant UHE cosmogenic neutrino flux~\cite{Muzio:2019leu}. Furthermore, the fit is also sensitive to magnetic fields~\cite{Wittkowski:2017okb,AlvesBatista:2024czs}.

Several cosmogenic neutrino estimates that can be found in the literature, in particular involving specific source populations, do not fit $\sigma(X_{\rm max})$, where $X_{\rm max}$ is the atmospheric depth at maximum shower development, which enables a higher flux at UHE. The RMS data can only be ignored if one assumes drastic deviations in hadronic models, i.e., when introducing exotic hadronic models~\cite{Albrecht_2022}.

The conservative interpretation of the Auger fits leads to a low cosmogenic UHE neutrino level. Possible caveats to this conclusion do not stem from hadronic model uncertainties, on which there is wide consensus, but rather on astrophysical models. 
Indeed, the existing particle source and propagation models are simplistic and based on various assumptions. In particular, there is a strong dependence on the maximum distance of UHECR sources (see Fig. 10 in Ref.~\cite{Alves_2019}), which could increase the flux by 2 orders of magnitude.

In spite of these uncertainties, because it is a guaranteed flux, given the existence of UHECRs and their extragalactic origin, most projects aim to position their 10-year diffuse limit to a level where the standard source parameters for UHECR could be ruled out \cite{2022NatRP...4..697G}: at around 
\begin{equation}
\label{eq:cosmostdmin}
E_\nu^2\Phi_{\rm cosmo, std, min}\,\sim 2\times 10^{-10}\,{\rm GeV}\,{\rm cm}^{-2}\,{\rm s}^{-1}\,{\rm sr}^{-1}\, \mbox{for } E_\nu \sim 3\times 10^{17}-3\times 10^{18}\,{\rm eV}\ .
\end{equation}

But detection could happen before we reach that limit, opening the path to actual UHE neutrino astronomy. Indeed, astrophysical UHE neutrinos could be produced abundantly directly at the sources~\cite{Unger_2015}, where accelerated UHECRs interact with the radiative and baryonic environments. 

For calculating the number of events we distinguish between the instantaneous and the day-averaged sensitivities of the detectors. This paper focuses on two classes of instruments as discussed previously - wide FoV with shallower sensitivity and deeper sensitivity with a narrow FoV. A good way to represent the effective areas in this case is \emph{acceptance}, defined as the effective area times the FoV in steradians. We show this product of all-flavor average effective area and the day-averaged FoV for GRAND, IceCube-Gen2 Radio, RNO-G, and BEACON in Figure~\ref{fig:avg_effectiveArea} (\emph{top panel}). We see that as one would intuitively understand the acceptance for both GRAND and IceCube-Gen2 Radio is nearly comparable in the energy range $10^8\ {\rm GeV} - 10^{11}\ {\rm GeV}$, even though GRAND has a much deeper sensitivity and IceCube-Gen2 Radio a wider FoV. The acceptance shown in the figure will be used in calculations of diffuse flux events in Section.~\ref{section:diffuse_point_source} and also in Section~\ref{subsec:deppandnarrow}. The instantaneous effective areas that are used to compute the day-averaged effective areas have been computed by the collaborations, see~\cite{GRAND:2018iaj} for GRAND,~\cite{Gen2_TDR} for IceCube-Gen2 Radio,~\cite{RNO-G:2020rmc} for RNO-G, and~\cite{BEACON:2025qcq} for BEACON. We show the instantaneous effective areas for the most sensitive band for each experiment in the \emph{bottom} panel of Figure~\ref{fig:avg_effectiveArea}.

\subsubsection{Promising diffuse astrophysical fluxes} 

A sample of astrophysical diffuse flux predictions for various source populations has been compiled in Figure~2 of \cite{valera2023}, and we propose a slightly updated version in Figure~\ref{fig:diffuse}. 
Although this astrophysical flux is not guaranteed, as UHECR accelerators could be  transparent to interactions, carving into these lines would strongly constrain source parameters.

In Figure~\ref{fig:diffuse}, the astrophysical fluxes are indicated in blue dotted lines with the following labeling and corresponding references: 
\begin{description}
   \item [radio-loud AGNs \cite{Murase_2014}] maximum allowed astrophysical neutrino flux allowed by the current IceCube limit, from broad line region and dust torus model, when fitting the counterpart cosmogenic neutrino flux to the experimental limits.
    
    \item [pulsars \cite{PhysRevD.90.103005}] newly-born pulsars following a uniform source evolution history.

    \item [clusters-1  \cite{Fang:2017zjf}] jets of radio-loud AGN embedded in galaxy clusters, within a grand-unified multi-messenger mode, fitting the UHECR Auger and PeV neutrino IceCube data. This flux also comprises a cosmogenic component.
    
    \item [clusters-2  \cite{Kotera09}]  AGN embedded in a cluster of galaxy with a magnetized cool core with field $B=30\,\mu$G and a mixed UHECR composition.
    
\end{description}

\subsubsection{Benchmark flux of the effective nucleus-survival} 

Interestingly, these source populations reach their maximum UHE neutrino flux around a generic astrophysically motivated line: the effective nucleus-survival line obtained by Murase and Beacom (solid blue in Figure~\ref{fig:diffuse}). This line corresponds to the neutrino flux level produced inside the sources by iron UHECRs, assuming an effective photodisintegration optical depth of 1 \cite{Murase_2010}:
\begin{equation}
E_\nu^2\Phi_{\rm MB}\,\sim 2.5\times 10^{-9}\,(f_z/3)\,(A/56)^{-0.21}\,{\rm GeV}\,{\rm cm}^{-2}\,{\rm s}^{-1}\,{\rm sr}^{-1}\, \mbox{ for } E_\nu\sim 10^{17-19}\,{\rm eV}\ ,
\end{equation}
with $f_z$ being the source population evolution factor (equal to 0.6 for uniform source evolution history, and 3 for strong evolution history), and $A$ the nucleus mass number. In Figure~\ref{fig:diffuse}, $f_z=3$ and $A=56$, denoting a conservative scenario. 

The nucleus-survival line can be considered as the extension to heavy nuclei of the nucleon-survival \cite{WaxmanBahcall98} line. One could be as bold as to consider it as the new ``Waxman-Bahcall goal" for experiments. 

\subsubsection{The IceCube extension to UHE energies} 
Another interesting line to place in the most important plots for diffuse fluxes is an extension of the IceCube measured flux. The extrapolation of the muon neutrino power-law in $E_\nu^{-2.37}$ is shown with its error band in navy blue, in Figure~\ref{fig:diffuse}:
\begin{equation}
\label{eq:icextrapol}
E_\nu^2\Phi_{\rm IC, extrapol.} \sim 10^{-8}\,(E_\nu/10^{16}\,{\rm eV})^{-2.37}\,{\rm GeV}\,{\rm cm}^{-2}\,{\rm s}^{-1}\,{\rm sr}^{-1}\, \mbox{for } E_\nu\sim 10^{17-18}\,{\rm eV}\ , 
\end{equation}

Contrarily to the nucleus-survival line, this extension does not stem from a theoretical model, but is a natural place where a diffuse UHE neutrino flux could appear, if the energy budget was equally distributed in various energy ranges. 

There is no strong argument to support that neutrino sources at UHE and HE would be the same. It is highly possible that they are different and that the HE neutrino sources run out of energy before reaching UHE. In such a case, a non detection would be an important measurement, setting constraints on the physics of these (yet unknown) IceCube sources. \\

Figure~\ref{fig:diffuse} summarizes all these possible diffuse neutrino flux levels, and overlays the projected experimental differential sensitivities (pink lines). It is interesting to notice that most of the astrophysical scenarios pile up around the IceCube extrapolation. As noted in \cite{valera2023}, detection or source constraints prospects look rather promising over the next decade. 

Any detector that will reach a sensitivity below 
$E_\nu^2\Phi_{\rm IC, extrapol.}$ or $E_\nu^2\Phi_{\rm MB}$
will be in position to either detect, or at least strongly constrain source models. Whether they can do UHE neutrino astronomy requires to assess additional performances.

\subsection{What if a diffuse neutrino flux is detected? Lessons from IceCube}\label{section:diffuse_point_source}
The detection of a diffuse neutrino flux is not the end goal of the field of UHE neutrinos. In order to extract the next layer of information from these messengers, one would need to relate them to a source. The lesson from IceCube, is that even with a large number of events detected, and even though neutrinos travel undeflected by any magnetic fields, identifying their sources is not a straightforward endeavor, without excellent angular resolution, and it has been argued that the sub-degree resolution is necessary to test the majority of promising models for the diffuse neutrino flux~\cite{2016PhRvD..94j3006M}. Two major astrophysical difficulties are the large spacetime density of source populations which seemingly isotropizes their distribution in the sky, and the absence of cosmic horizon for neutrinos that can hence stem from the dawn of the Universe. Most of all, IceCube is limited in this search by its angular resolution, which has a majority of events reconstructed with angular resolution.
A significant improvement on these performances, in particular using machine learning techniques, have revealed a clustering of events originating from the Galactic plane \cite{IceCube_Science_2023}. Besides, the planned upgrade to IceCube~\cite{Ishihara:2019aao} will also increase its pointing abilities.

At UHE, studies have indeed shown that angular resolution will be needed on top of sensitivity, in order to pin-point sources in a diffuse flux. Steady sources with density (at $z=0$) $n_{\rm s}\sim 10^{-7}-10^{-5}\,{\rm Mpc}^{-3}$ not evolving over time ($n_{\rm s}\sim 10^{-9}-10^{-7}\,{\rm Mpc}^{-3}$ for a population density evolution following the star formation rate) can be identified as an event excess in the sky with $5\sigma$ significance only with the detection of $100-1000$ neutrinos and sub-degree angular resolution~\cite{Fang:2016hop}. 

\begin{figure}[!t]
\centering
\includegraphics[width=0.8\columnwidth]{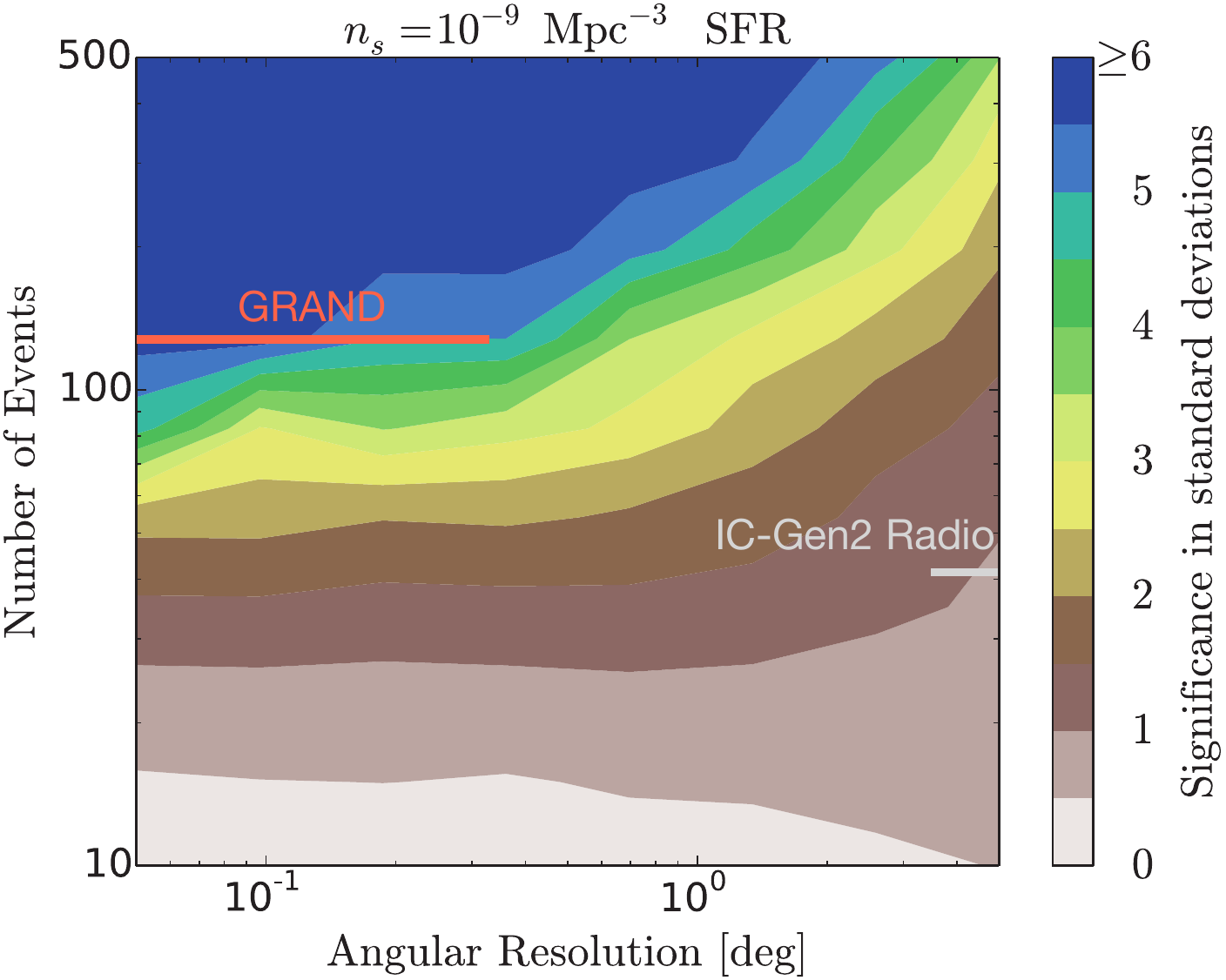}
\caption{Significance of detection of point sources, within a diffuse UHE neutrinos flux, by experiments with given angular resolutions and number of detected events. Here, we present the specific case of a source population density $n_{\rm s}= 10^{-9}\,{\rm Mpc}^{-3}$, following the star formation rate evolution, up to redshift $z=6$. With this source number density, $\sim 100$ events and $\sim 0.1^\circ$ angular resolution are needed to reach a $4\sigma$ detection of point sources within a diffuse flux. The angular resolution of GRAND is taken from~\cite{DecoenePhD20} while that for IceCube-Gen2 Radio from~\cite{2022NatRP...4..697G}. The observation time assumed is $10$ years for both the detectors. (Adapted from \cite{Fang:2016hop}.)
}
\label{fig:point_source}
\end{figure}

As an illustration, Figure~\ref{fig:point_source} (adapted from \cite{Fang:2016hop}) shows the significance of detection of point sources by neutrino experiments, as a function of their angular resolution and number of events detected, for the specific case of a source population density $n_{\rm s}= 10^{-9}\,{\rm Mpc}^{-3}$, following the star formation rate evolution, up to redshift 6. 
The color coding corresponds to the confidence level to reject an isotropic background using the statistical method from \cite{Fang_2016}. 

The maroon and gray lines correspond respectively to GRAND and IceCube-Gen2 Radio expectations for event arrival direction reconstruction \cite{Decoene_2023,Barwick:2021wU,bouma2023directionreconstructioniniceradio} and for number of neutrinos detected, assuming a Murase-Beacom flux~\cite{Murase_2010} of $E_\nu^2\Phi_{\rm MB}=2.5\times 10^{-9}\,{\rm GeV}\,{\rm cm}^{-2}\,{\rm s}^{-1}\,{\rm sr}^{-1}$ over $E_\nu\sim 10^{17-19}\,$eV, for an observation time of 10 years. 

In Figure~\ref{fig:point_source}, a daily averaged field of view coverage of $f_{\rm dFoV} = 100\%$ is used. We discussed in \cite{Fang_2016} that fewer events are required in the field of view if $f_{\rm dFoV}$ is smaller. In general, to reach the same significance level of detection, more events will
be needed if sources have a larger source number density, or if more sources are distributed at large
distances.

\section{Populations of transient sources and their detection strategies}\label{section:transient}

As argued in e.g., Ref.~\cite{2022NatRP...4..697G}, anisotropy, source-density, energetics, and magnetic-structure arguments strongly challenge steady-source scenarios for UHECRs with light composition \cite{2019FrASS...6...23B, Abreu13_18, Anchordoqui:2018qom, 2016ApJ...832L..17F, Palladino_2020}. Following these arguments, powerful transients, with their large amount of energy injected over short timescales, can be considered as the most promising sources when it comes to producing UHE neutrinos. Section~\ref{section:diffuse_point_source} also demonstrates the difficulty of identifying steady sources among a nearly isotropic background of neutrinos. In this respect, transient sources offer the additional timing information for coincident searches. 

In the following, we recall the characteristics of the most powerful transient populations and assess their detectability with different types of instruments. 

\subsection{Transient population characteristics}
Transient populations deemed capable of producing UHE neutrinos have been compiled and described in recent studies and reviews \cite{2019ARNPS..69..477M, Guepin17, 2022NatRP...4..697G}. 

To compare their fluences to instrumental performances, transients can be categorized in two types: \emph{short} and \emph{long} transients. The fluence of the former can be compared to instantaneous instrumental sensitivities (the source has to stay in the instantaneous FoV and present a stable fluence over that time -- so this depends on the instrument, but correspond typically to durations of less than 1\,h). 
For longer transients (from 1\,h to months), a day-averaged sensitivity should be used.

Table~\ref{tab:transients} is based on the references cited above, and lists the two principal characteristics of transients necessary to roughly assess their detectability: their typical minimum and maximum bright luminosity durations ($t_{\rm s}$) at corresponding isotropic equivalent bolometric luminosities ($L_{\rm s}$), global real population rates and outflow Lorentz factors ($\Gamma_{
\rm s}$).

\begin{table}
\caption{Ranges of typical properties of different categories of transient sources (duration $\log_{10} t_{\rm s}$, isotropic equivalent bolometric luminosity $\log_{10} L_{\rm s}$, real population rate $\log_{10} {\cal R}_{\rm s}$, outflow Lorentz factor $\Gamma_{\rm s}$). Short bursts: stay in the instantaneous field of view of the instrument (typically $\sim 1\,$hr). 
Long bursts: any longer transients. For beamed sources, the apparent (observed) rate can be computed from ${\cal R}_{\rm s}$ using ${\cal R}_{\rm app} = (\theta^2/2) {\cal R}_{\rm s}$, where $\theta \sim 1/\Gamma_s$. Note: Tidal Disruption Events (TDE), Superluminous Supernovae (SLSNe), Hypernovae (HNe), Supernovae Type Ibc (SNIbc), Binary Neutron Star (BNS), Binary Black Hole (BBH), Binary White Dwarf (BWD), Low-Luminosity Gamma-Ray Bursts (LLGRB), High-Luminosity Gamma-Ray Bursts (HLGRB).\\}
\centering
\resizebox{\textwidth}{!}{%
\begin{tabular}{lccccc}
\hline
\textbf{Source} & Long/Short & $\log_{10} t_{\rm s}$ (s) & $\log_{10} L_{\rm s}$ (erg s$^{-1}$) & $\log_{10} {\cal R}_{\rm s}$ (Gpc$^{-3}$ yr$^{-1}$) & $\Gamma_{\rm s}$ \\
\hline
TDE jetted & Short/Long & $2-7$ & $47-48$ & $-1.5$ to $3$ & $100$ \\
TDE non-jetted & Short/Long & $2-7$ & $42-45$ & $2-3$ & $1$ \\
Blazar Flares & Short/Long & $2-6$ & $44-50$ & $0-2$ & $3-100$ \\
Pulsars (fast-rotating) & Long & $3-8$ & $44-45$ & $3.4 - 5.6$ & $1$ \\
Magnetars & Short & $3-4$ & $41 - 46$  & $3.4 - 5.6$ & $1$ \\
& Long & $4-5$ & $41 - 48$   & $3.4 - 5.6$ & $1$ \\
SLSNe, HNe & Long/Short & $5-7$ & $43-45$ & $3.5 - 4.5$ & $1$ \\
SNIbc & Long/Short & $5-7$ & $42-43$ & $4.5 - 4.6$ & $1$ \\
BNS Mergers & Long/Short & $3-4$ & $46-48$ & $1-3.2$ & $1-100$ \\
BBH Mergers & Long/Short & $4-6.7$ & $43-46$ & $1.2-1.8$ & $1-100$ \\
BWD Mergers & Long/Short & $2-4$ & $44-46$ & $3-4$ & $1-10$ \\
LLGRB & Short & $3-4$ & $46-49$ & $2-3$ & $1-10$ \\
HLGRB & Short & $-3 - 2$ & $49-53$ & $2-3$ & $100-300$ \\
\hline
\end{tabular}
}
\label{tab:transients}
\end{table}

\begin{figure*}[!t]
\centering
\includegraphics[width=0.8\columnwidth]{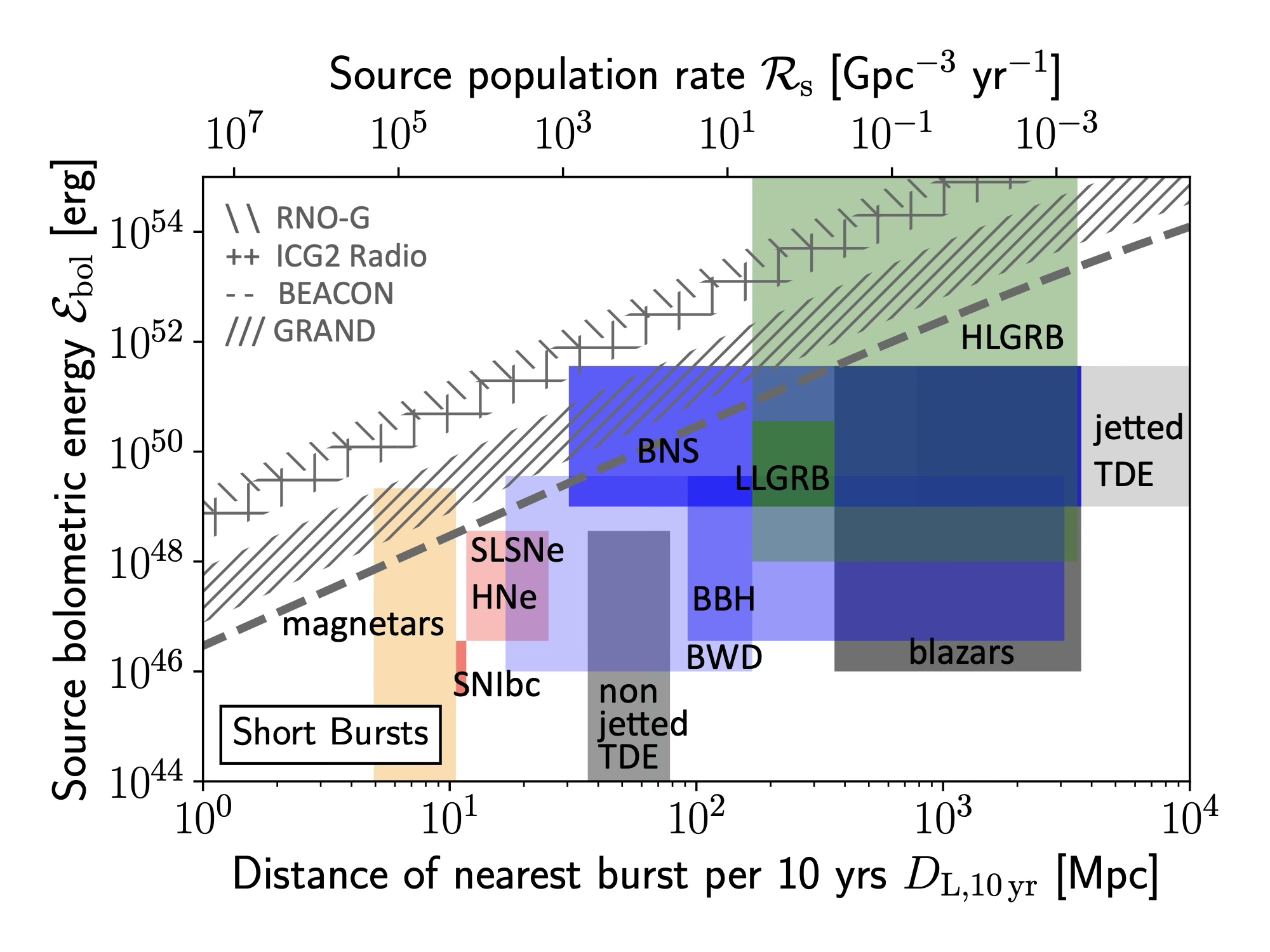}
\includegraphics[width=0.8\columnwidth]{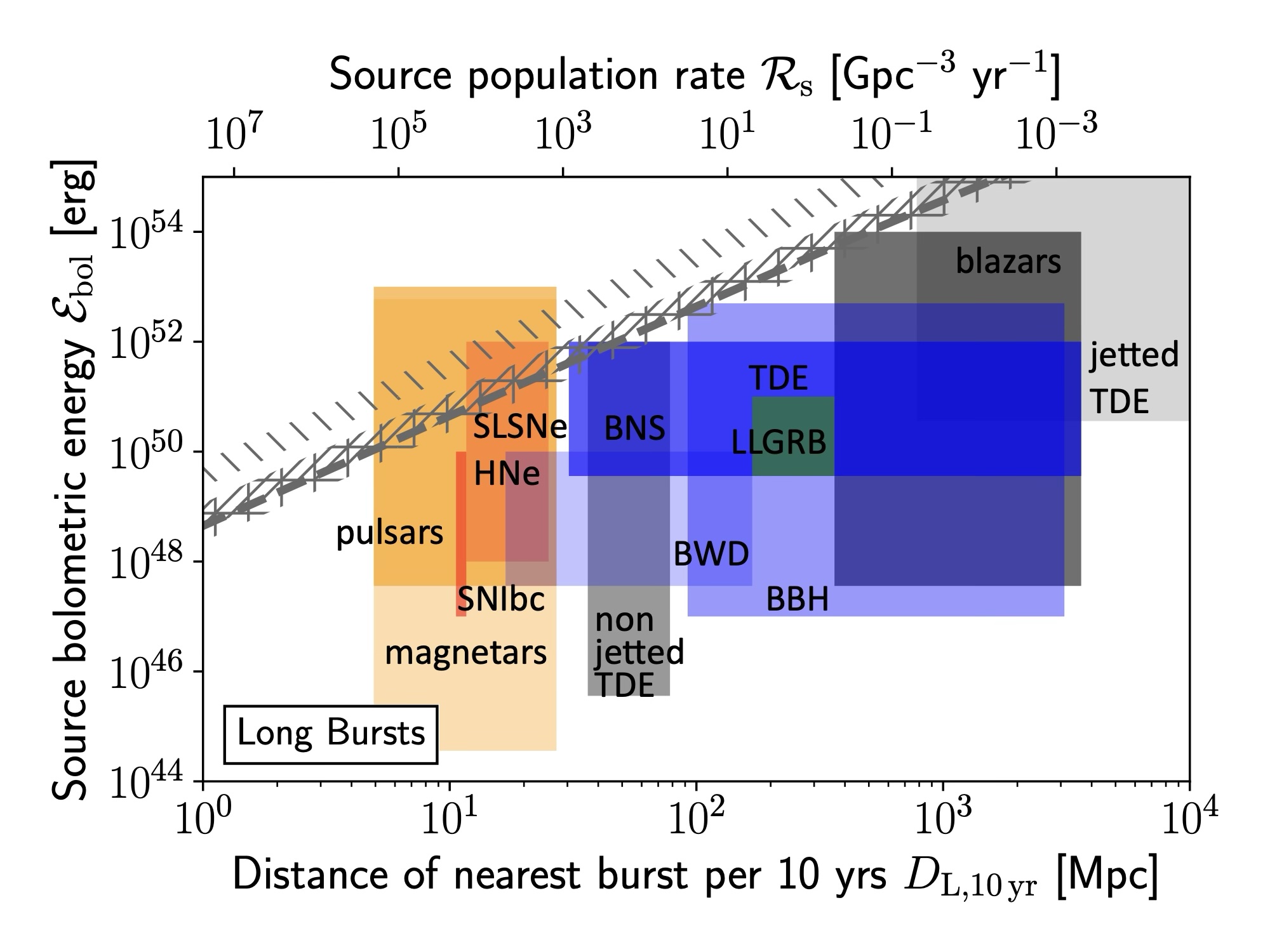}
\caption{Detection possibility of short (\emph{top}) and long (\emph{bottom}) transients within a given distance, within 10 years of observation. The regions spanned by each source populations in the ${\cal E}_{\rm bol}-D_{\rm L,10\,yr}$ (bolometric energy, typical distance for a 10\,yr observation) parameter space are presented, following Table~\ref{tab:transients}, assuming conservatively $f_\nu^{\rm UHE}=1$. Maximal experimental sensitivity bands (for $E_\nu\gtrsim 10^{17.5}\,$eV) are overlaid. The most promising short transients are distant and bright, while the long ones are local. Detection strategies will be different for both populations.}
\label{fig:EvsD}
\end{figure*}

\begin{figure*}[!t]
\centering
\includegraphics[width=0.8\columnwidth]{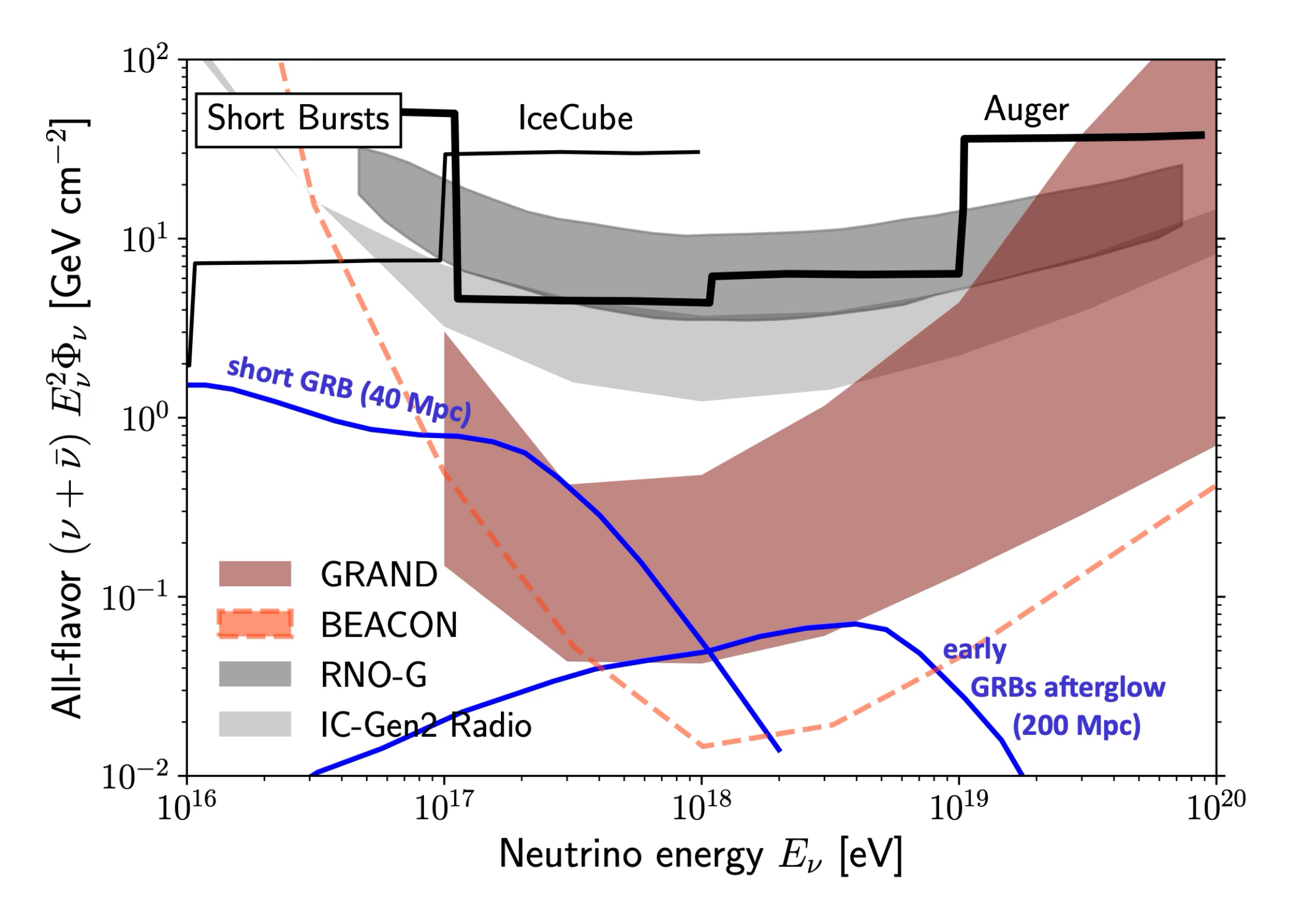}
\includegraphics[width=0.8\columnwidth]{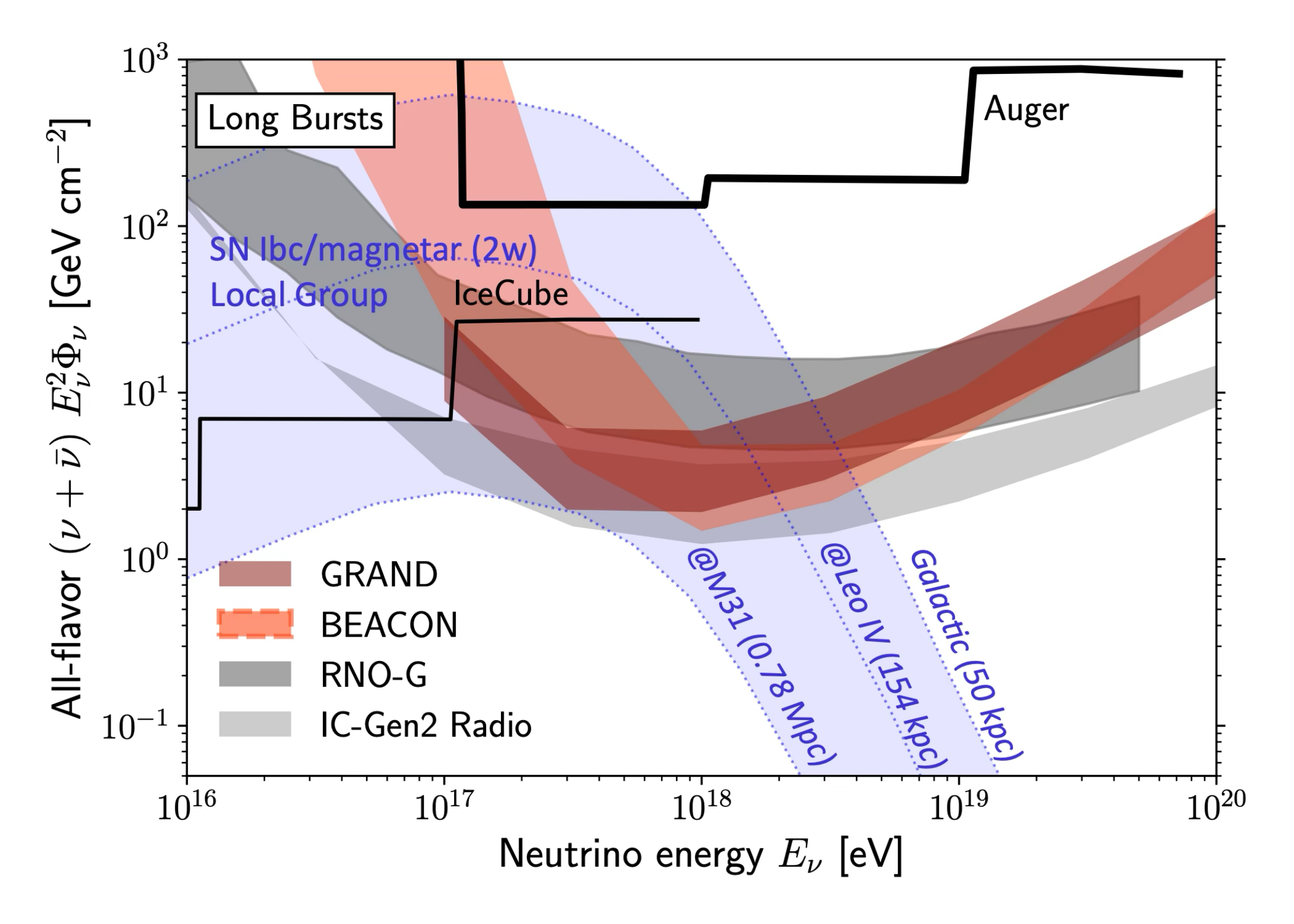}
\caption{All-flavor neutrino fluence sensitivities per decade in energy for an assumed $E_\nu^{-2}$ neutrino spectrum for various upcoming experiments, as indicated. For Auger and IceCube~\cite{2017ApJ...850L..35A}, the upper-limit sensitivity to neutrino transients at the location of GW170817A is shown. Overlaid are theoretical neutrino fluences for (\emph{top}) a short GRB/binary neutron star merger at 40\,Mpc \cite{Kimura:2017kan}, an early GRB afterglow at 200\,Mpc \cite{Murase:2007yt}, (\emph{bottom}) a SN Ibc supernova with magnetar 2 weeks after birth \cite{Murase:2009pg}, located at the indicated distances, inside the Galaxy, or galaxies of the Local Group (Leo IV and M31). Adapted from~\cite{2022NatRP...4..697G}.
}
\label{fig:fluence_sens}
\end{figure*}

The criteria for a transient source to be able to accelerate neutrinos to ultrahigh energies, and the corresponding fluence calculation are detailed in Ref.~\cite{Guepin17} and reviewed in \cite{2022NatRP...4..697G}. Here, we do not specify the neutrino production mechanism and assume that a fraction $f_\nu^{\rm UHE}$ of the isotropic equivalent bolometric energy of the outflow ${\cal E}_{\rm bol}=L_{\rm s}\,t_{\rm s}$ is channeled into UHE neutrino fluence:
\begin{equation}
\label{eq:flux}
E_\nu^2\phi_\nu  = \frac{(1+z)}{4 \pi D_L^2} f_\nu^{\rm UHE}\,{\cal E}_{\rm bol},
\end{equation}
with $z$ the source redshift and $D_{\rm L}$ the corresponding luminosity distance. 

For a given real source population rate ${\cal R}_{\rm s}$, the typical luminosity distance $D_{\rm L,T_{\rm obs}}$ for a burst occurring over an observation time $T_{\rm obs}$ can be estimated by 
\begin{equation}
D_{\rm L,T_{\rm obs}} = \left(\frac{3}{4\pi\,f_{\rm beam}\,{\cal R}_{\rm s}\,T_{\rm obs}}\right)^{1/3}\ ,
\end{equation}
with $f_{\rm beam}$ the beaming fraction, which can be related to the outflow beaming angle $\theta\sim 1/\Gamma_{\rm s}$ via $f_{\rm beam}\sim \theta^2/2$ \cite{Piran2005}. For beamed sources, the apparent (observed) rate can be computed from ${\cal R}_{\rm s}$ using ${\cal R}_{\rm app} = f_{\rm beam} {\cal R}_{\rm s}$.

For each source population, we estimate roughly, using the bounds quoted in Table~\ref{tab:transients}, the minimum and maximum bolometric energies ${\cal E}_{\rm bol}$ and typical burst distance $D_{\rm L, 10\, yr}$ for a $T_{\rm obs, 10\,{\rm yr}}=10$\,year observation time by 
\begin{eqnarray}
 {\cal E}_{\rm bol, min}&=&\frac{L_{\rm s, min}}{t_{\rm s,max}}\ \mbox{and}\ {\cal E}_{\rm bol, max}=\frac{L_{\rm s, max}}{t_{\rm s,min}}\
,\\
D_{\rm L,10\,{\rm yr}, min} &=& \left(\frac{6\,\Gamma_{\rm s,min}^2}{4\pi\,{\cal R}_{\rm s,max}\,T_{\rm obs, 10\,{\rm yr}}}\right)^{1/3}\ ,\\
D_{\rm L,10\,{\rm yr}, max} &=& \left(\frac{6\,\Gamma_{\rm s,max}^2}{4\pi\,{\cal R}_{\rm s,min}\,T_{\rm obs, 10\,{\rm yr}}}\right)^{1/3}\ ,
\end{eqnarray}
with $t_{\rm short,min} = \min[t_{\rm s},1\,{\rm hr}]$ for short sources and $t_{\rm long,min} = \max[t_{\rm s},1\,{\rm hr}]$, to account for sources that can be considered as both long or short sources, depending on their evolution (see Table~\ref{tab:transients}). 

A detailed examination of these short/long states, duration, corresponding luminosities, beaming factors and rates would necessitate a case-by-case population study, beyond the scope of this generic strategical paper. 

The regions spanned by each source populations in the ${\cal E}_{\rm bol}-D_{\rm L,10\,yr}$ parameter space are presented for short (left panel) and long (right panel) bursts in Figure~\ref{fig:EvsD}, assuming $f_{\nu}^{\rm UHE} = 1$. This last assumption sets a conservative upper limit to the source energy level reached. More realistic values span $f_{\nu}^{\rm UHE} \lesssim 0.1$.

\subsection{Detectability}
In order to assess whether source populations might have a chance to be detected by envisioned instruments, we overlay in Figure~\ref{fig:EvsD} the instrumental fluence sensitivities ${\cal S}_{\rm exp}$, i.e, the Feldman-Cousins upper limit~\cite{Feldman_1998} per decade in energy at 90\% C.L., assuming a power-law all-flavor neutrino spectrum $E_\nu^{-2}$, for no candidate events and null background. In practice, we equate Eq.~\eqref{eq:flux} with 
\begin{equation}
E_\nu^2 {\cal S}_{\rm exp}  = \frac{2.44\, N_\nu}{\Delta(\log_{10}E_{\rm \nu}) \ln(10)}\frac{E_\nu}{A_{\rm eff}}\ ,
\end{equation}
where $N_\nu=3$ accounts for the 3 neutrino flavors, $\Delta (\log_{10} E_{\rm \nu}) \,\ln(10)$  is the size of the logarithmic energy bin (here, one decade: $\Delta (\log_{10} E_{\rm \nu})=1 $) over which a power-law all-flavor neutrino spectrum $E_\nu^{-2}$ is integrated, given the all-flavor effective area $A_{\rm eff}$. The factor $2.44$ corresponds to the upper limit on the number of events detected, for the 90\% C.L. interval assuming a Poisson distribution of number of events with zero background~\cite{Feldman_1998}. We use the simulated differential neutrino energy, declination and right-ascension dependent effective areas $A_{\rm eff}$ directly provided by the collaborations of indicated instruments. 

The width of the shaded regions corresponds to the zenith angle dependency, at maximum experimental sensitivity (all experiments reach this at $E_\nu \gtrsim 10^{17.5}\,$eV). Sources with parameter-regions within or above the gray bands could be detectable by these experiments. 

From Figure~\ref{fig:EvsD}, it appears clearly that one will require different observation strategies for short and long sources. Indeed, the most promising short transients (even, and especially, taking into account a significantly lower $f_\nu^{\rm UHE}$) for UHE neutrino detection are the rare, hence distant, bright gamma-ray bursts. On the other hand, for longer transients, local sources such as pulsars, magnetars, or superluminous supernovae appear as promising candidates.

\subsection{Deep and narrow vs wide and shallow: instrumental sensitivities}

Experimental fluence sensitivities are presented as a function of energy for short and long bursts in Figure~\ref{fig:fluence_sens}, in a figure adapted from \cite{2022NatRP...4..697G}, for BEACON (1000 stations)~\cite{Wissel:2020sec,Zeolla:ARENA2024}, GRAND (200k antennas)~\cite{Kotera:2024iyk}, IceCube-Gen2 Radio \cite{Gen2_TDR} and RNO-G \cite{RNO-G:2020rmc}. 

The left-hand panel presents the instantaneous sensitivity for a short transient source located in the instrument field of view. The blue line indicates the neutrino fluence of a short GRB associated with a binary neutron star merger located at 40~Mpc for an on-axis viewing, following the model of \cite{Kimura:2017kan}.

The right-hand panel presents the day-averaged sensitivity: for each declination, the experimental effective area is averaged over the right ascensions. The detectability of bursts with duration longer than the time for a source to remain in the field of view of narrow instruments should be assessed with these day-averaged sensitivities. 

For GRAND, the maroon band encompasses zenith angles $86^{\circ} \leq \theta \leq 93^{\circ}$ for short transients and declination $0^{\circ} \leq \delta \leq 45^{\circ}$ for long transients, assuming a single GRAND array at latitude $42^{\circ}$ North~\cite{GRAND:2018iaj}.
For IceCube-Gen2 Radio, the band covers zenith angles $0^{\circ} \leq \theta \leq 60^{\circ}$. As IceCube-Gen2 Radio has a homogeneous effective area over all right ascensions, the instantaneous and daily averaged sensitivities are equal. For BEACON, the maximum instantaneous or right-ascension averaged sensitivity over all declinations is presented. For RNO-G, the instantaneous band encompasses zenith angles $46^\circ-83^\circ$, and the day-averaged band corresponds to the average over right ascensions for declinations $-15^\circ$ to $55^\circ$.

Note that the balloon mission PUEO~\cite{2021JInst..16P8035A} has, in principle, the deepest instantaneous sensitivity among all the planned instruments, albeit at more than 10 times higher energy threshold. We do not quote its limit here, because of this energy regime, and because taking into account its operation duration limited to a month is beyond the scope of this paper.

Overlaid are theoretical neutrino fluences of single sources, for promising populations, as identified in Figure~\ref{fig:EvsD}. Among short bursts (left): a binary neutron star merger associated to a short GRB at 40\,Mpc \cite{Kimura:2017kan}, an early GRB afterglow at 200\,Mpc \cite{Murase:2007yt}. For long transients: a supernova Ibc harboring a fast-rotating magnetar, 2 weeks after birth, located at indicated distances as indicated, inside the Galaxy, or galaxies of the Local Group (Leo IV and M31). For the latter, we rescaled the model of \cite{Murase:2009pg}, assuming a more conservative magnetar dipole magnetic field strength of $B=10^{15}\,$G and 1\,ms rotation period at birth. 

The two types of instruments discussed in the introduction appear in Figure~\ref{fig:fluence_sens}: while they both align on the day-averaged sensitivity, the narrow FoV detectors compensate with deep instantaneous sensitivities, while the wide FoV detectors lose in instantaneous sensitivity. 
\subsection{Deep and narrow instruments: more powerful to detect rare and bright short bursts}
\label{subsec:deppandnarrow}

\begin{figure*}[!t]
\centering
\includegraphics[width=\textwidth]{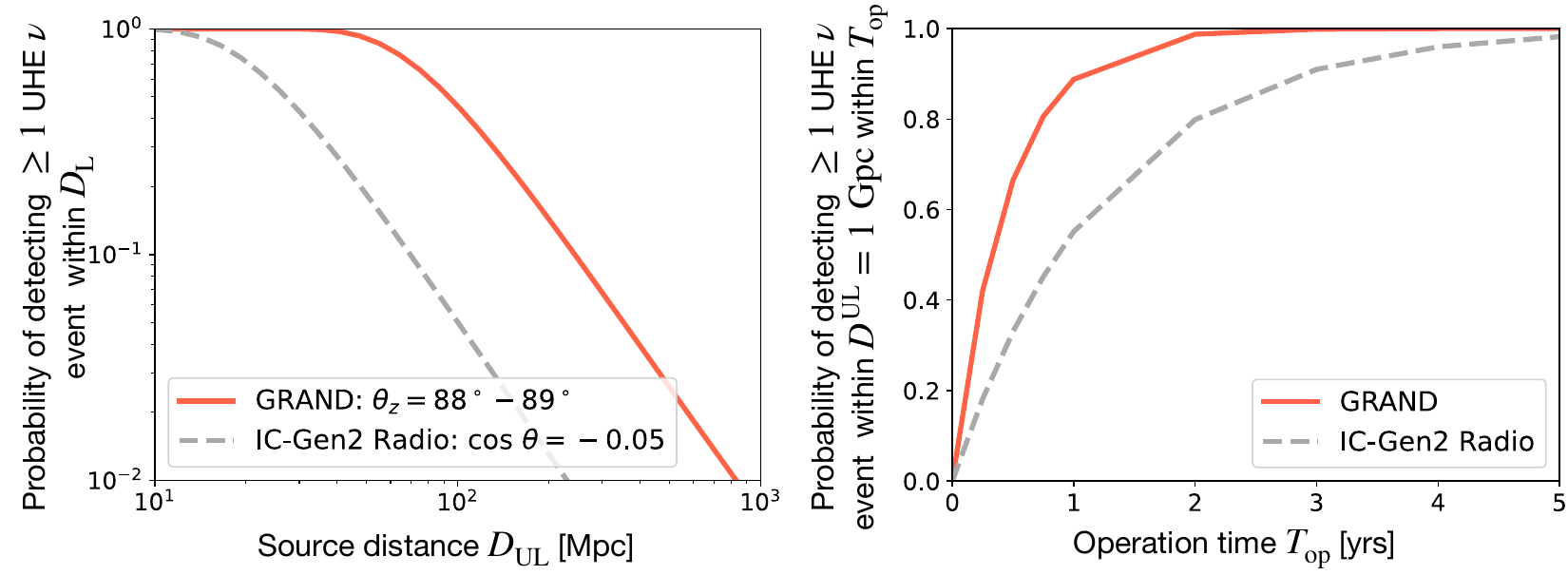}
\caption{\emph{Left:} The probability ($P_{n\geq 1} (D_{\rm L})$, Eq.~\ref{eq:totprob}) of detecting at least one UHE neutrino event from a typical BNS merger within a distance $D_{\rm L}$ assuming the most sensitive instantaneous field of view for GRAND (solid) and IceCube-Gen2 Radio (dashed). \emph{Right:} The probability ($P(D^{\rm UL})$, Eq.~\ref{eq:qtop}) of detecting one UHE neutrino event from typical BNS merger events given a distance horizon $D^{\rm UL} = 1 {\rm Gpc}$ within $T_{\rm op}$ years of operation for GRAND and IceCube-Gen2 Radio. In this case, the all-flavor average effective area is used, as shown in Figure~\ref{fig:avg_effectiveArea}.
} 
\label{fig:detection_proba}
\end{figure*}

Figure~\ref{fig:EvsD} points towards rare and bright sources (HLGRBs in particular) as most promising short transients as detectable UHE neutrino producers. Targeting these sources requires to integrate over large volumes of the Universe to collect as many events as possible that are observed by electromagnetic (EM) instruments, and stack the fluxes at their location. Similar stacking strategies with defined time windows as in IceCube stacked transient searches (e.g. \cite{Abbasi_2024} or specifically for gamma-ray bursts, \cite{Abbasi_2022}) could be performed, although with a largely reduced background, as no atmospheric neutrinos are expected at these energies. Note that this follow-up strategy also requires decent angular resolution for pointing. 

Intuitively, increasing the FoV is an effect in distance squared, while increasing the instantaneous sensitivity depth amounts to increasing the volume, hence an effect in distance cubed. For this reason, at equal diffuse flux sensitivity, deep and narrow instruments will perform better than wide and shallow ones, for short burst detection. 

Let us now try to quantify this. It is known that the probability of detecting at least one UHE neutrino event within a luminosity distance $D_{\rm L}$ is given by
\be
\label{eq:totprob}
P_{n\geq 1} (D_{\rm L}) = \frac{1}{\Omega_{\rm norm}} \int d\Omega\ p_{n \geq 1} \big(\theta_z, D_{\rm L} \big)\,,
\ee
where the Poissonian probability to detect at least one UHE neutrino event is given as, $p_{n \geq 1} \big(\theta_z, D_{\rm L} \big) = 1 - {\rm exp} \big( -N_\nu (\theta_z, D_{\rm L}) \big)$~\cite{Kimura:2017kan,Kheirandish_2023,Mukhopadhyay:2023niv,Mukhopadhyay:2024lwq} and this is integrated over the solid angle $\Omega$. The normalization $\Omega_{\rm norm}$ is the solid angle over which the probability is averaged ($\Omega_{\rm norm} = 4 \pi$ for all-sky). 
The total number of neutrino events at a given declination band is given by $N_\nu (\theta_z, D_{\rm L})$. We choose the most sensitive bands for GRAND and IceCube-Gen2 Radio and show the results for $P_{n\geq 1} (D_{\rm L})$ in the left panel of Figure~\ref{fig:detection_proba}. For $D_{\rm L} \lesssim 10$ Mpc both GRAND and IceCube-Gen2 Radio have close to $100\%$ probability of detecting an UHE neutrino event from a source. However, owing to the narrow and deep sensitivity of GRAND as a result of the higher instantaneous effective areas, the probability of detecting an UHE neutrino event remains close to $1$ for distances close to $100$ Mpc.

The above probability is averaged over the angle. It is also natural to introduce the similar probability for the cumulative number of events within a given distance horizon $D_{\rm L} = D^{\rm UL}$, considering the line-of-sight integral. Note that we consider the probability for the number of events, which is different from that for the number of sources~\cite{2016PhRvD..94j3006M,Yoshida22}. We essentially follow the formalism outlined in Refs.~\cite{Mukhopadhyay:2023niv,Mukhopadhyay:2024lwq}. 
Given an upper limit for a luminosity distance $D^{\rm UL}$, which is determined by other observational information such as gravitational waves or electromagnetic telescopes, the probability to detect at least one UHE neutrino event can be given by
\begin{align}
\label{eq:qtop}
P(D^{\rm UL}) &= 1 - {\rm exp}\bigg( -\mathcal{N} \big( D^{\rm UL} \big) \bigg) \nonumber\,,\\
\mathcal{N} \big( D^{\rm UL} \big) = \int_{0}^{D^{\rm UL}} d (&d_{\rm com}) \frac{T_{\rm op}}{\big( 1+z \big)} R_{\rm app}\big( z \big) d_{\rm com}^2 \tilde{N}_\nu (d_{\rm com})\,,
\end{align}
where $T_{\rm op}$ is the operation time of the UHE neutrino detector, $d_{\rm com}$ is the comoving distance, $R_{\rm app} (z)$ is the apparent redshift ($z$) dependent rate of a given source. The total number of events is given by $\tilde{N}_\nu = \int_{E_\nu^{\rm min}}^{E_\nu^{\rm max}} dE_\nu\ \phi_\nu\ \tilde{A}_{\rm eff} (E_\nu)$, where $\phi_\nu$ is the UHE neutrino fluence from a typical source, $E_\nu$ is the neutrino energy in the observer frame (see Ref.~\cite{Mukhopadhyay:2024lwq} for details), and $\tilde{A}_{\rm eff} (E_\nu)$ is the all-flavor average effective-area times the day-averaged field-of-view ($\Delta \Omega$) in steradians. Note that the $4\pi$ appearing as a result of the spherical integration cancels with the $4\pi$ appearing as the denominator for the fraction of sky area covered with respect to the field-of-view, that is, $\Delta \Omega/4\pi$. The results for $\tilde{A}_{\rm eff} (E_\nu)$ from GRAND, IceCube-Gen2 Radio, and RNO-G are shown in Figure~\ref{fig:avg_effectiveArea}. 

The choice of $D^{\rm UL}$ depends on science cases. For core-collapse supernovae, we may use the distance horizon set by optical telescopes. For compact object mergers, this distance can be set by gravitational wave detectors. For a source like a BNS merger, one can compute the probability to detect at least one UHE neutrino event using Eq.~\eqref{eq:qtop} given $D^{\rm UL}$, where $R_{\rm app}(z)$ and $\phi_\nu$ are the apparent redshift dependent rate of and the UHE neutrino fluence from the BNS merger respectively. In Figure~\ref{fig:detection_proba} right panel, we show the results for $D^{\rm UL} = 1$ Gpc. It is evident that in this case GRAND can detect an UHE neutrino event in roughly $\sim 3$ years of operation time, while IceCube-Gen2 Radio has $\sim 90\%$ probability of seeing an UHE neutrino event within similar timescales of operation. Note that the distance $D^{\rm UL}$ can be inferred from a GW- or EM-triggered search for UHE neutrinos from BNS mergers~\cite{Mukhopadhyay:2023niv,Mukhopadhyay:2024lwq}, in which case $T_{\rm op}$ would be the joint operation timescale of the UHE neutrino and GW or EM detectors respectively. On another note, for upcoming experiments a time-dependent effective area can help with performing more detailed searches using time-dependent neutrino light curves from models.

\section{Nearby transients: sample of host galaxies in the Local Universe}\label{section:nearby}

\begin{figure*}[!t]
\centering
\includegraphics[width=\textwidth]{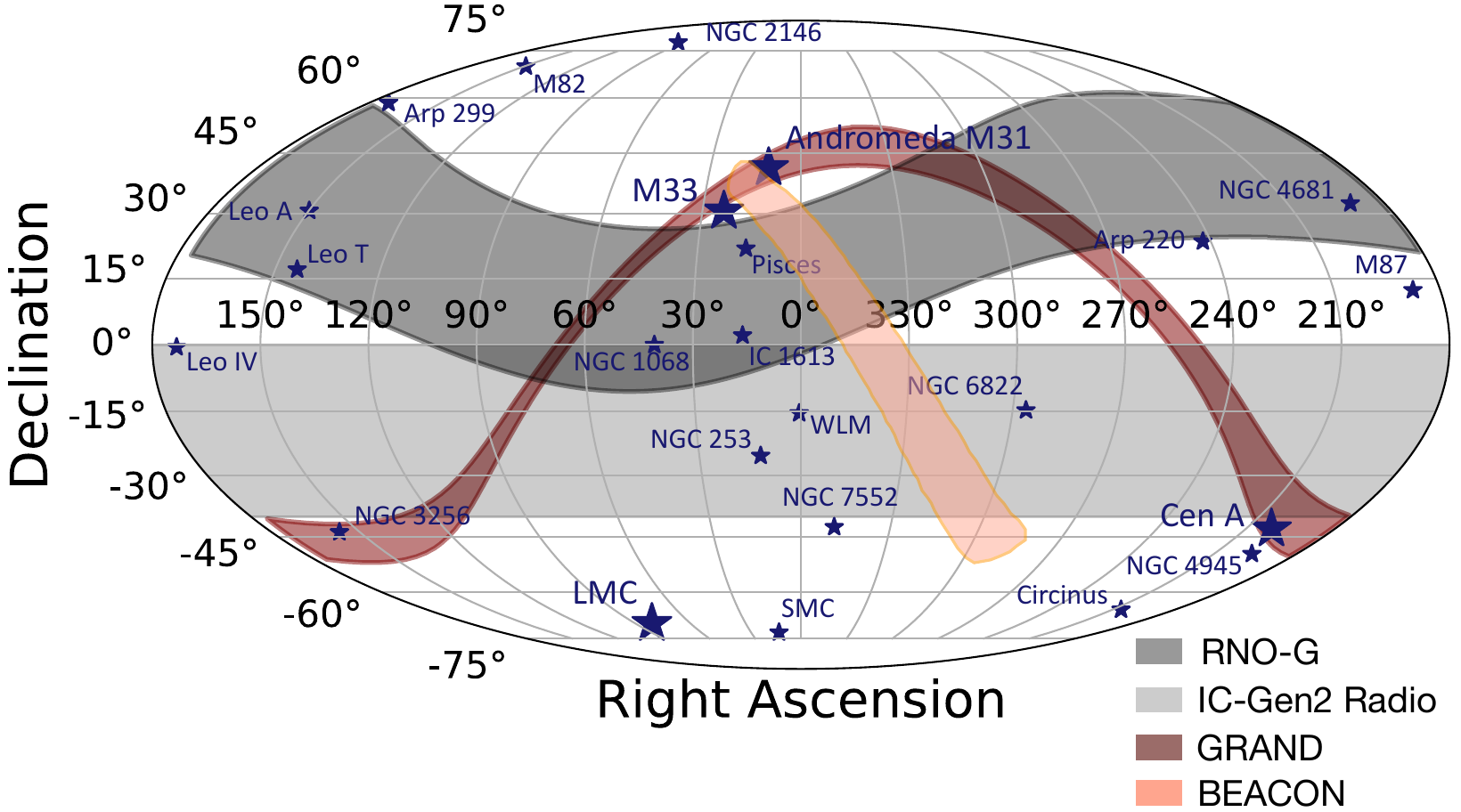}
\caption{Local galaxies (see Table~\ref{tab:local_group}) and instantaneous FoV of various experiments. See Table 1 in~\cite{2022NatRP...4..697G} for additional details on the various experiments.}
\label{fig:instFOV_skymap}
\end{figure*}

\begin{figure*}[!t]
\centering
\includegraphics[width=\columnwidth]{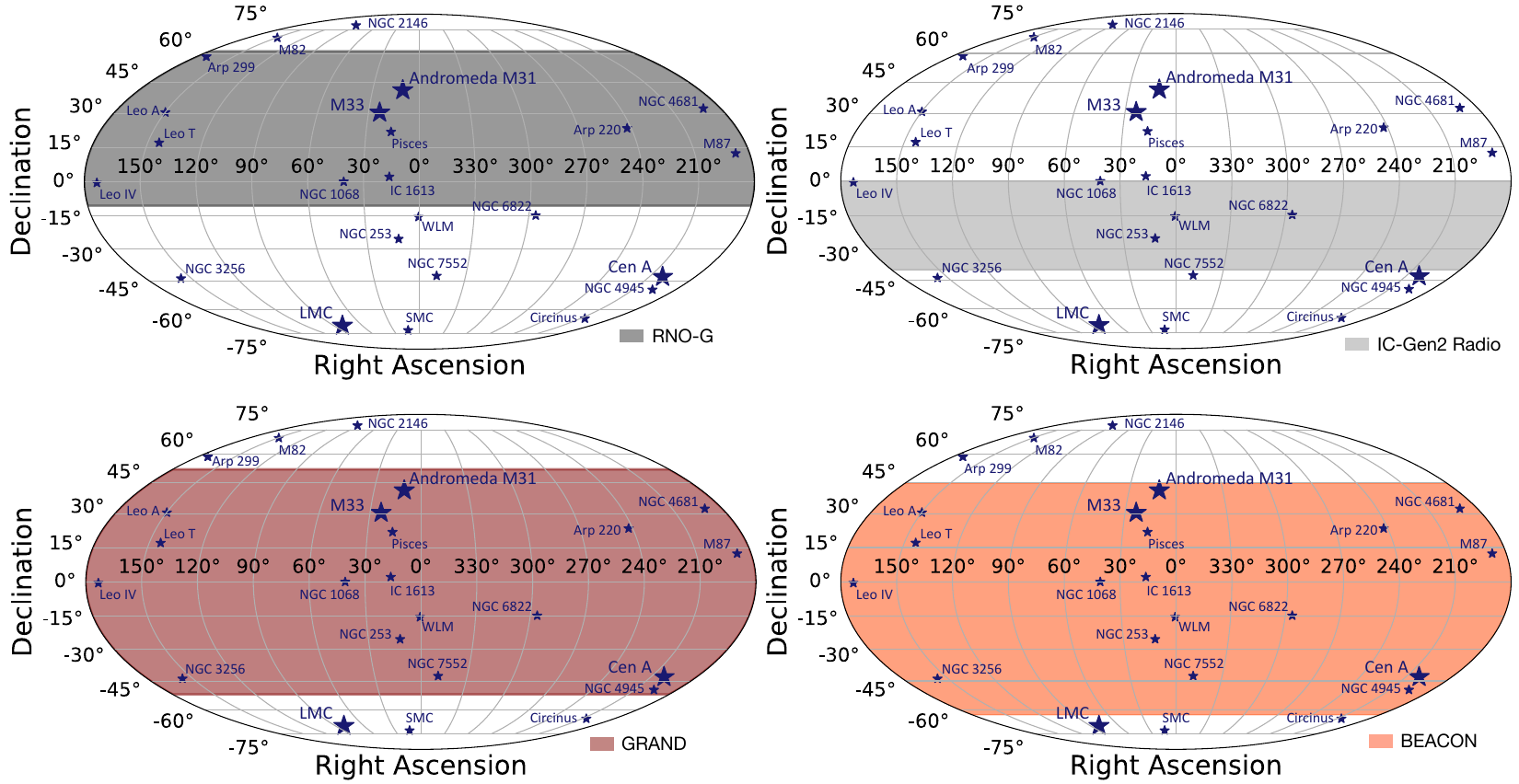}
\caption{Local galaxies (see Table~\ref{tab:local_group}) and daily (day-averaged) FoV of various experiments. See Table 1 in~\cite{2022NatRP...4..697G} for additional details on the various experiments.
}
\label{fig:avgFOV_skymap}
\end{figure*}

Figure~\ref{fig:EvsD} revealed that some of the most promising sources to be detected in UHE neutrinos were long transients in our local Universe -- spanning from the Local Group to few tens of Mpc. These present a reasonable combination of luminosity, closeness, and population rate. To detect these events, the largest possible day-averaged FoV is favored. For instruments that are not located at the Poles, the Earth's rotation advantageously increases the overall day-FoV to reach up to 80\% sky-coverage. The location of the various detectors on the globe is hence of prime importance to obtain the best collective sky-coverage. 

In order to illustrate this point, we present in Figures~\ref{fig:instFOV_skymap} and~\ref{fig:avgFOV_skymap} the instantaneous and day-averaged FoV of a few instruments, together with the positions of a sample of host galaxies in our local Universe, focusing in particular on the Local Group and on nearby starburst galaxies, that could harbor the most promising nearby transients for UHE neutrino detection. Instruments might want to consider having these galaxies in their day-average FoV, if they have a choice on the orientation of their instrument and/or on the location on Earth.

Table~\ref{tab:local_group} lists the host galaxies in our Local Group that are represented in the skymaps of Figures~\ref{fig:instFOV_skymap} and~\ref{fig:avgFOV_skymap}. This adhoc sample consists of i) the closest galaxies in our Local Group presenting possible evidence of active star formation, or with low metallicity, i.e, with a large population of old massive stars. ii) Significant starburst galaxies within $\sim 80\,$Mpc. iii) M87, a giant elliptical galaxy located in the Virgo Cluster and one of the most massive galaxies in the Local Universe. M87 is also an AGN and presents evidence of ongoing star formation in its outer regions, although its rate is lower than in typical starburst galaxies.
iv) Centaurus A (Cen~A), a radio-loud AGN with powerful relativistic jets. It is the closest AGN with a supermassive black hole at its center that is actively accreting matter. It has long been considered as a promising candidate for producing ultrahigh-energy (UHE) neutrinos through hadronic interactions of cosmic rays accelerated in its jet and surrounding environments \cite{Cuoco:2007aa,Kachelriess:2008qx,Mbarek:2024nvv}.

We caution that this list is not exhaustive, and a more systematic census of the galaxies in the Local Universe should be undertaken in order to provide a reliable catalog for potential source surveillance. Other compilations of nearby sources with high core-collapse supernova rates, for multi-messenger searches (and neutrino production at lower energies) can be found in \cite{Nakamura_2016, Kheirandish_2023}.

In the following paragraphs, we provide more details on the nature, rates and localisation of potential nearby sources. 

\subsection{Pulsars, magnetars, and supernovae rates} 
Magnetars and pulsars stand out as most promising sources in the long-burst population panel of Figure~\ref{fig:EvsD}. 
The formation of these objects is linked to supernovae in environments with high star formation rates. According to \cite{Beniamini_2019}, the birth rate of magnetars is estimated to be $\sim 2-20/{\rm kyr}/{\rm galaxy}$, depending on the specifics of the stellar environment. For the Local Group, this implies an expected magnetar birth rate of $\sim 0.16-1.6/{\rm yr}$, or one magnetar every 1 to 6 years.

Due to their rotational and magnetic energy reservoirs, pulsars and magnetars are capable of accelerating cosmic rays to UHE~\cite{Bednarek97,Blasi00,Arons03,Fang12}, a process that can lead to the production of UHE neutrinos as in the right panel of Figure~\ref{fig:fluence_sens}~\cite{Murase:2009pg,Fang14,PhysRevD.90.103005,Fang15,Fang:2017tla,Carpio:2020wzg,Mukhopadhyay:2024ehs,Mukhopadhyay:2025tvz} (and also to associated GW signatures~\cite{K11}). Not all pulsars and magnetars are expected to generate UHE neutrinos, but estimates suggest that a fraction of $\sim 1-10\%$ of the total population may have the right properties to contribute to the UHE neutrino flux \cite{Murase:2009pg,Fang12,Fang14,Mukhopadhyay:2024ehs}. This would result in an occurrence rate of a UHE neutrino producer in the Local Group on the order of one event every $10-600$\,years.

This rate is to be linked to that of superluminous supernovae (SLSNe) and Type Ibc supernovae (SN Ibc), that could be powered by powerful pulsars and magnetars at birth \cite{Kasen10, Quimby12, Dessart12, KPO13, Kashiyama_2016, Metzger_2015,Rodriguez:2024txg}. These associated events would provide an EM signature to follow-up on.

\subsection{Host galaxies in the Local Group}

The Local Group, a collection of more than 80 galaxies located within approximately 3 Mpc, consists of 3 large galaxies (the Milky Way, Andromeda or M31, and M33) and dwarf galaxies (see~\cite{Karachentsev_2004} for a full catalog). A handful of these galaxies exhibit high star formation rates. Even though many dwarf and irregular galaxies in the Local Group have low ongoing star formation rates, their supernova occurrence probability could be boosted due to factors such as past bursts of star formation, the presence of older stellar populations, or interactions with other galaxies \cite{Tolstoy2009}. 

Dwarf galaxies in the Local Group often show episodic star formation, with starbursts driven by external factors such as tidal interactions or internal gas dynamics \cite{2004AJ....127.1531H,2005astro.ph..6430D,Weisz2011,Geha_2024}. 
Besides, these galaxies exhibit low metallicity, leading to a top-heavy Initial Mass Function (IMF), that produces a higher fraction of massive stars \cite{Kauffmann2003}. Supernova rates are largely tied to the population of massive stars \cite{Maoz2014}, and in particular the rate of SN Ibc \cite{Habergham_2010,Anderson13}. 

\subsection{Starburst galaxies in the Local Universe}

Starburst galaxies, such as M82, NGC 253, and NGC 1068, Arp 299, exhibit extremely high star formation rates, making them favorable environments for the formation of engine-driven supernovae including low-luminosity GRBs \cite{Zhang:2017moz,Zhang:2018agl}, powerful pulsars, and magnetars~\cite{Arons03,Murase:2006mm}. Moreover, starburst galaxies have long been considered as ideal sites for cosmic ray acceleration, e.g., in the shock waves of supernovae or stellar winds, and of subsequent high-energy neutrino production via interactions with the dense interstellar medium~\cite{Loeb:2006tw,Lacki:2010vs,Senno:2015tra}.

\begin{table}
\caption{Properties of a sample of host galaxies in the Local Group and nearby starburst galaxies, that could harbor transient sources capable of producing observable UHE neutrinos. Properties include galaxy type, distance $D_{\rm L}$, celestial coordinates, mass, star formation rate (SFR), and metallicity ($Z$).}
\centering
\resizebox{\textwidth}{!}{%
\begin{tabular}{lcccccccc}
\hline  
\textbf{Galaxy} & {Type} & \begin{tabular}{@{}c@{}} $D_{\rm L}$ \\(kpc) \end{tabular}  &  Right Ascension & Declination  & \begin{tabular}{@{}c@{}} Mass \\($M_\odot$)\end{tabular}& \begin{tabular}{@{}c@{}} SFR\\ $(M_\odot/{\rm yr})$\end{tabular} & \begin{tabular}{@{}c@{}} $Z$\\($Z_\odot$)\end{tabular} & Comment\\
\hline
Andromeda (M31) & Spiral & 765 & 00h 42m 44.3s & +41d 16m 9s & $1.5\times 10^{12}$ & $0.35$ & 0.5 & Largest galaxy in the Group. \\
Triangulum (M33) & Spiral & 97 & 01h 33m 50.02s & +30d 39m 36.7s & $5\times 10^{10}$ & $0.45$ & 0.2 & 3rd largest galaxy. \\
LMC & Spiral & 5 & 05h 23m 34s & -69d 45.4m & $1\times 10^{10}$ & $0.2$ & 0.5& 4th largest galaxy, 2nd closest. \\
WLM & Irregular & 93 & 00h 01m 58.1s & -15d 27m 39s & $4.3\times 10^8$ & $0.02$ & 0.1 & At least one recent supernova. \\
SMC & Irregular & 62 & 00h 52m 44.8s & -72d 49m 43s & $7\times 10^9$ & $0.03$ & 0.2 & Moderate SFR. \\
Pisces & Irregular & 769 & 01h 03m 55.0s & +21d 53m 06s & $3\times 10^9$ & $0.05$ & 0.05 & Low mass, irregular galaxy. \\
IC1613 & Irregular & 73 & 01h 04m 47.8s & +02d 07m 04s & $2.4\times 10^9$ & $0.02$ & 0.03 & Recent supernovae observed. \\
Leo A & Irregular & 79 & 09h 59m 26.4s & +30d 44m 47s & $4\times 10^8$ & $0.01$ & 0.02 & Active ongoing SF. \\
NGC 6822 & Irregular & 500 & 19h 44m 56.6s & -14d 47m 21s & $5\times 10^9$ & $0.03$ & 0.03 & Active ongoing SF. \\
Leo IV & Dwarf Sph. & 154 & 11h 32m 57s & -00d 32m 00s & $2\times 10^8$ & $0.0002$ & 0.0002 & Recent star-formation history. \\
Leo T & Dwarf Sph. & 420 & 09h 34m 53.4s & +17d 03m 05s & $1\times 10^8$ & $0.0001$ & 0.0001 & SF only recently.\\
Circinus Galaxy & Spiral & 1290 & 13h 43m 44.04s & -65d 19m 46s & $2\times 10^{10}$ & $2.5-6$ & 1.2  & Type-II Seyfert, closest to Milky Way. \\
\hline
M82 & Spiral & 3500 & 09h 55m 52.6s & +69d 40m 46s & $7.5\times 10^{10}$ & $10-20$ & 0.2  & Active starburst, interaction with M81. \\
NGC 253 & Spiral & 3500 & 00h 47m 33s & -25d 17m 17s & $7.5\times 10^{10}$ & $3-5$ & 0.3  & Active starburst, several recent supernovae. \\
NGC 4945 & Spiral & 3800 & 13h 09m 2.2s & -49d 28m 06s & $1.5\times 10^{10}$ & $1 - 2$ & 0.8 & Nearby, evidence of SF.\\
NGC 4631 & Spiral & 5097 & 12h 42m 08s & +32d 32m 03s & $3\times 10^{10}$ & $1.5$ & 0.3  & Active starburst. \\
NGC 1068 & Spiral & 14400 & 02h 42m 40.77s & -00d 00m 47.84s & $1 \times 10^9$ & 3.2 & 1.056 & Type-II Seyfert galaxy, active starburst. \\
NGC 7552 & Spiral & 17200 & 23h 15m 18s & -42d 33m 39s & $4\times 10^{10}$ & $5$ & 0.3  & Active starburst. \\
NGC 2146 & Spiral & 18000 & 06h 18m 38.4s & +78d 21m 36s & $2\times 10^{11}$ & $5 - 15$ & 1.0 & Active starburst.\\
NGC 3256 & Peculiar & 37400 & 10h 27m 45.3s & -43d 51m 56s & $3\times 10^{10}$ & $10-20$ & 0.3 & Active starburst: galaxy merger. \\
Arp 299/NGC 3690 & Peculiar & 47700 & 11h 28m 30.6s & +58d 33m 38s & $2\times 10^{11}$ & $50 - 100$ &  1.2  & Extreme starburst: galaxy merger. \\
Arp 220 & Peculiar & 78000 & 16h 09m 17s & +23d 30m 45s & $3.5\times 10^{11}$ & $200$ &  0.5  & Extreme starburst: galaxy merger. \\
\hline
M87 & Elliptical & 16400 & 12h 30m 49.4s & +12d 23m 28s & $1.2\times 10^{12}$ & $0.1$ & 0.5 & Largest galaxy in the Virgo Cluster. \\
\hline
Cen A & Lenticular & 4000 & 13h 25m 28s & -43d 1m 9s & $2\times 10^{11}$ & $0.5$ & 0.3 & Radio-loud AGN. \\
\hline
\end{tabular}
}
\label{tab:local_group}
\end{table}

\section{Conclusions, Perspectives}
\label{sec:concl}

The new multi-messenger and time domain landscape leads us to update and reconsider our science goals towards our understanding of the extremely energetic sources in the Universe. This consequently inspires us to review our  strategies when designing next-generation experiments at ultrahigh energies. In this paper, we have summarized, cross-analysed, and interpreted some key multi-messenger advances in this perspective. We summarize our main conclusions and propose a set of observation strategies to increase our chances to observe UHE neutrinos, with a relevant scientific outcome.  

\subsection{What diffuse UHE neutrino sensitivities to reach? What angular resolution?}

Any detector that will reach a sensitivity below the IceCube extrapolated flux $\Phi_{\rm IC, extrapol.}\sim 10^{-8}\,(E_\nu/10^{16}\,{\rm eV})^{-2.37}\,{\rm GeV}\,{\rm cm}^{-2}\,{\rm s}^{-1}\,{\rm sr}^{-1}$ or the nucleus-survival benchmark flux $E_\nu^2\Phi_{\rm MB}\sim 2.5\times 10^{-9}\,(f_z/3)\,(A/56)^{-0.21}\,{\rm GeV}\,{\rm cm}^{-2}\,{\rm s}^{-1}\,{\rm sr}^{-1}$ between $E_\nu=10^{17-18}\,$eV typically, will be in a position to either detect, or at least strongly constrain source models. 

To go further and enable UHE neutrino astronomy, instruments will need to reach sub-degree angular resolution. With all future instruments aligning at $\Phi_{\rm IC extrapol.}$ (Eq.~\ref{eq:icextrapol}) or better, at $\Phi_{\rm cosmo,std,min}$ (Eq.~\ref{eq:cosmostdmin}), the difference in achievable science by measuring the diffuse flux will be driven primarily by angular resolution.  

\subsection{Astrophysically motivated strategy for UHE neutrino detection}

Theoretically, sources of UHE neutrinos are expected to be extremely energetic, hence bright in EM (at some frequency) or gravitational waves (GW). Their population should be already known to us. In such cases, our bolometric energy-distance parameter scan for known transients points towards two types of promising sources:

\begin{enumerate}
\item Short rare and bright transients. In that case, follow-up searches on EM or GW signals and stacking at these locations appears as the best strategy to increase UHE neutrino detection probability. {\it Deep instruments with sub-degree angular resolution are better suited for this purpose.}

\item Nearby serendipitous sources (with occurrence rate 1 every $10-600$\, yrs). These would be long types of transients, likely well-identified and observed in all wavelengths. Hence {\it both deep \& narrow and wide \& shallow instruments would be in position to detect such events}. The variety of instrumental performances would provide complementary information in terms of angular resolution (for a better statistical significance of the event coincidence), time coverage with possible time variabilities (wide instantaneous FoV instruments being more performing if the source is in their FoV, though the sensitivity has to be high enough to measure flux variations), energy resolution, flavor measurement. 

\end{enumerate}

Finally, serendipitous yet unknown sources could exist, that are huge emitters of UHE neutrino though being either invisible counterparts in EM and GW, or are too rare to have been detected yet. For such serendipitous objects, a large instantaneous field of view, so a wide and shallow instrument, provides a better survey, and the best chance of catching a source.

\subsection{What can we do to improve UHE neutrino detectability and the associated scientific output?}

In light of the discussions and conclusions of this work, we list below some aspects for future instruments to consider in their design, and what theory and phenomenology can provide to help. Although the community that has traditionally focused on improving the experimental sensitivity to the diffuse UHE neutrino flux, the latest evolutions in the field of multi-messenger astronomy leads us to re-think this strategy. 

\begin{enumerate}
\item Improve the instantaneous sensitivity even at the cost of reduced instantaneous FoV (deep \& narrow)
\item Improve the angular resolution down to sub-degree.
\item Build a catalog of sources in the Local Universe that instruments should have in their day-averaged FoV. 
\item Follow-up these catalogued sources. For most UHE instruments that do not point and track source positions in the sky naturally, this requires to develop a designated data-taking/observation mode. 
\item For narrow instruments: Widen the instantaneous FoV along right ascension, to increase sensitivity for long bursts. 
\item Coordinate and optimize the location of detectors on the globe for best collective daily sky coverage. 
\end{enumerate}

Comment on Point 2: In-air detectors naturally have a sub-degree angular resolution thanks, in particular, to the lever arm provided by the large extension of the air-shower signals. In-ice experiments are limited in their angular reconstruction by the difficulty to reconstruct the polarization of their radio signal. As in the case of IceCube, machine learning techniques provide promising tools to improve on this limitation at data analysis level.  

Comment on Point 4: such a follow-up data-taking mode would run in parallel to the usual triggering modes. Techniques could be developed to narrow the radio beam via antenna phasing as in BEACON~\cite{Wissel:2020sec,Southall:2022yil}, or point via radio interferometry \cite{Schoorlemmer:2020low,Schluter:2021egm}.

Comment on Point 5: For in-air radio experiments, where the instantaneous FoV is limited by the intrinsic detection technique (neutrinos arrive from a limited slice of Earth where it interacts and from which the secondary tau particle escapes), the major handle for improvement resides in topography. Mountains can widen the FoV band by a few degrees \cite{GRAND:2018iaj,Decoene:2019eux}. 

To achieve simultaneously wide field-of-view, good angular resolution, and deep sensitivity, an ideal solution could be to deploy several “deep \& narrow” experiments at different sites worldwide (ideally at mid-latitudes and spanning different longitudes).

Finally, we note that a wide and shallow instrument would be fully complementary to a deep and narrow search, hence the ideal situation would be to build both types of instruments. These would, furthermore, rely on different experimental methods and hence improve the overall quality of data, increase the quantity of information, and the chance of detection.

\begin{acknowledgments}
The authors thank Jaime Alvarez-Muñiz, Aurélien Benoit-Lévy, Valentin Decoene, Claire Guépin, Ryan Krebs, Foteini Oikonomou, Mauricio Bustamante, and Tanguy Pierog for very fruitful discussions.
K.\,K. acknowledges support from the Fulbright-France program, the CNRS Programme Blanc MITI ("GRAND" 2023.1 268448; France), and the CNRS Programme AMORCE ("GRAND" 258540; France). M.\,M. and K.M. are supported by NSF Grant No. AST-2108466. M.\,M. also acknowledges support from the Institute for Gravitation and the Cosmos (IGC) Postdoctoral Fellowship. The work of K.\,M. is supported by the NSF Grant Nos. AST-2108467 and AST-2308021, and JSPS KAKENHI Grant 
No.~20H05852. R.\,A.B. is supported by the Agence Nationale de la Recherche (ANR), project ANR-23-CPJ1-0103-01. O.\,M. is  supported by the ANR (ANR-21-CE31-0025) and the DFG (Projektnummer
490843803). S.\,W. and A.\,Z. are supported by NSF Award 2033500 and S.\,W. is further supported by NASA Awards 80NSSC24K1780, 80NSSC22K1519, and 80NSSC21M0116. 
\end{acknowledgments}
\bibliography{NatRev}

@article{Abbasi_2024,
doi = {10.3847/1538-4357/ad18d6},
url = {https://dx.doi.org/10.3847/1538-4357/ad18d6},
year = {2024},
month = {mar},
publisher = {The American Astronomical Society},
volume = {964},
number = {1},
pages = {40},
author = {Abbasi, R. and Ackermann, M. and Adams, J. and IceCube Collaboration},
title = {Search for Continuous and Transient Neutrino Emission Associated with IceCube Highest-energy Tracks: An 11 yr Analysis},
journal = {The Astrophysical Journal},
}

@article{Abbasi_2022,
   title={Searches for Neutrinos from Gamma-Ray Bursts Using the IceCube Neutrino Observatory},
   volume={939},
   ISSN={1538-4357},
   url={http://dx.doi.org/10.3847/1538-4357/ac9785},
   DOI={10.3847/1538-4357/ac9785},
   number={2},
   journal={The Astrophysical Journal},
   publisher={American Astronomical Society},
   author={Abbasi, R. and Ackermann, M. and Adams, J. and  others},
   year={2022},
   month=nov, pages={116} }

@article{Murase_2014,
   title={Diffuse neutrino intensity from the inner jets of active galactic nuclei: Impacts of external photon fields and the blazar sequence},
   volume={90},
   ISSN={1550-2368},
   url={http://dx.doi.org/10.1103/PhysRevD.90.023007},
   DOI={10.1103/physrevd.90.023007},
   number={2},
   journal={Physical Review D},
   publisher={American Physical Society (APS)},
   author={Murase, Kohta and Inoue, Yoshiyuki and Dermer, Charles D.},
   year={2014},
   month=jul }

@article{muzio2024peterscycleendcosmic,
    author = "Muzio, Marco Stein and Anchordoqui, Luis A. and Unger, Michael",
    title = "{Peters cycle at the end of the cosmic ray spectrum?}",
    eprint = "2309.16518",
    archivePrefix = "arXiv",
    primaryClass = "astro-ph.HE",
    doi = "10.1103/PhysRevD.109.023006",
    journal = "Phys. Rev. D",
    volume = "109",
    number = "2",
    pages = "023006",
    year = "2024"
}

@article{Unger_2015,
   title={Origin of the ankle in the ultrahigh energy cosmic ray spectrum, and of the extragalactic protons below it},
   volume={92},
   ISSN={1550-2368},
   url={http://dx.doi.org/10.1103/PhysRevD.92.123001},
   DOI={10.1103/physrevd.92.123001},
   number={12},
   journal={Physical Review D},
   publisher={American Physical Society (APS)},
   author={Unger, Michael and Farrar, Glennys R. and Anchordoqui, Luis A.},
   year={2015},
   month=dec }

@article{Southall:2022yil,
    author = "Southall, Dan and others",
    title = "{Design and initial performance of the prototype for the BEACON instrument for detection of ultrahigh energy particles}",
    eprint = "2206.09660",
    archivePrefix = "arXiv",
    primaryClass = "astro-ph.IM",
    doi = "10.1016/j.nima.2022.167889",
    journal = "Nucl. Instrum. Meth. A",
    volume = "1048",
    pages = "167889",
    year = "2023"
}

@article{Schoorlemmer:2020low,
    author = "Schoorlemmer, Harm and Carvalho, Washington R.",
    title = "{Radio interferometry applied to the observation of cosmic-ray induced extensive air showers}",
    doi = "10.1140/epjc/s10052-021-09925-9",
    journal = "Eur. Phys. J. C",
    volume = "81",
    number = "12",
    pages = "1120",
    year = "2021"
}

@article{Schluter:2021egm,
    author = {Schl\"uter, Felix and Huege, Tim},
    title = "{Expected performance of air-shower measurements with the radio-interferometric technique}",
    doi = "10.1088/1748-0221/16/07/P07048",
    journal = "JINST",
    volume = "16",
    number = "07",
    pages = "P07048",
    year = "2021"
}

@article{Albrecht_2022,
   title={The Muon Puzzle in cosmic-ray induced air showers and its connection to the Large Hadron Collider},
   volume={367},
   ISSN={1572-946X},
   url={http://dx.doi.org/10.1007/s10509-022-04054-5},
   DOI={10.1007/s10509-022-04054-5},
   number={3},
   journal={Astrophysics and Space Science},
   publisher={Springer Science and Business Media LLC},
   author={Albrecht, Johannes and Cazon, Lorenzo and Dembinski, Hans and Fedynitch, Anatoli and Kampert, Karl-Heinz and Pierog, Tanguy and Rhode, Wolfgang and Soldin, Dennis and Spaan, Bernhard and Ulrich, Ralf and Unger, Michael},
   year={2022},
   month=mar }

@article{Anderson13,
	author = {{Anderson, J. P.} and {Soto, M.}},
	title = {On the multiplicity of supernovae within host galaxies},
	DOI= "10.1051/0004-6361/201220600",
	url= "https://doi.org/10.1051/0004-6361/201220600",
	journal = {A\&A},
	year = 2013,
	volume = 550,
	pages = "A69",
	month = "",
}

@article{Habergham_2010,
doi = {10.1088/0004-637X/717/1/342},
url = {https://dx.doi.org/10.1088/0004-637X/717/1/342},
year = {2010},
month = {jun},
publisher = {The American Astronomical Society},
volume = {717},
number = {1},
pages = {342},
author = {S. M. Habergham and J. P. Anderson and P. A. James},
title = {TYPE I bc SUPERNOVAE IN DISTURBED GALAXIES: EVIDENCE FOR A TOP-HEAVY INITIAL MASS FUNCTION},
journal = {The Astrophysical Journal}
}

@article{Karachentsev_2004,
doi = {10.1086/382905},
url = {https://dx.doi.org/10.1086/382905},
year = {2004},
month = {apr},
publisher = {},
volume = {127},
number = {4},
pages = {2031},
author = {Igor D. Karachentsev and Valentina E. Karachentseva and Walter K. Huchtmeier and Dmitry I. Makarov},
title = {A Catalog of Neighboring Galaxies},
journal = {The Astronomical Journal},
}

@article{Kheirandish_2023,
   title={Detecting High-energy Neutrino Minibursts from Local Supernovae with Multiple Neutrino Observatories},
   volume={956},
   ISSN={2041-8213},
   url={http://dx.doi.org/10.3847/2041-8213/acf84f},
   DOI={10.3847/2041-8213/acf84f},
   number={1},
   journal={The Astrophysical Journal Letters},
   publisher={American Astronomical Society},
   author={Kheirandish, Ali and Murase, Kohta},
   year={2023},
   month=oct, pages={L8} }

@article{Nakamura_2016,
   title={Multimessenger signals of long-term core-collapse supernova simulations: synergetic observation strategies},
   volume={461},
   ISSN={1365-2966},
   url={http://dx.doi.org/10.1093/mnras/stw1453},
   DOI={10.1093/mnras/stw1453},
   number={3},
   journal={Monthly Notices of the Royal Astronomical Society},
   publisher={Oxford University Press (OUP)},
   author={Nakamura, Ko and Horiuchi, Shunsaku and Tanaka, Masaomi and Hayama, Kazuhiro and Takiwaki, Tomoya and Kotake, Kei},
   year={2016},
   month=jun, pages={3296–3313} }

@article{Zhang:2017moz,
    author = "Zhang, B. Theodore and Murase, Kohta and Kimura, Shigeo S. and Horiuchi, Shunsaku and M\'esz\'aros, Peter",
    title = "{Low-luminosity gamma-ray bursts as the sources of ultrahigh-energy cosmic ray nuclei}",
    eprint = "1712.09984",
    archivePrefix = "arXiv",
    primaryClass = "astro-ph.HE",
    doi = "10.1103/PhysRevD.97.083010",
    journal = "Phys. Rev. D",
    volume = "97",
    number = "8",
    pages = "083010",
    year = "2018"
}

@article{Zhang:2018agl,
    author = "Zhang, B. Theodore and Murase, Kohta",
    title = "{Ultrahigh-energy cosmic-ray nuclei and neutrinos from engine-driven supernovae}",
    eprint = "1812.10289",
    archivePrefix = "arXiv",
    primaryClass = "astro-ph.HE",
    doi = "10.1103/PhysRevD.100.103004",
    journal = "Phys. Rev. D",
    volume = "100",
    number = "10",
    pages = "103004",
    year = "2019"
}

@ARTICLE{2013Sci...342E...1I,
       author = {{IceCube Collaboration}},
        title = "{Evidence for High-Energy Extraterrestrial Neutrinos at the IceCube Detector}",
      journal = {Science},
         year = 2013,
        month = nov,
       volume = {342},
       number = {6161},
          eid = {1242856},
        pages = {1242856},
          doi = {10.1126/science.1242856},
archivePrefix = {arXiv},
       eprint = {1311.5238},
 primaryClass = {astro-ph.HE},
       adsurl = {https://ui.adsabs.harvard.edu/abs/2013Sci...342E...1I}
}

@ARTICLE{Maoz2014,
       author = {{Maoz}, Dan and {Mannucci}, Filippo and {Nelemans}, Gijs},
        title = "{Observational Clues to the Progenitors of Type Ia Supernovae}",
      journal = {Annual Reviews of Astronomy and Astrophysics},
         year = 2014,
        month = aug,
       volume = {52},
        pages = {107-170},
          doi = {10.1146/annurev-astro-082812-141031},
archivePrefix = {arXiv},
       eprint = {1312.0628},
 primaryClass = {astro-ph.CO},
       adsurl = {https://ui.adsabs.harvard.edu/abs/2014ARA&A..52..107M}
}

@article{Tolstoy2009,
  author = {E. Tolstoy and others},
  title = {Star Formation Histories of Local Group Dwarf Galaxies},
  journal = {Annual Review of Astronomy and Astrophysics},
  volume = {47},
  year = {2009},
  pages = {371-425},
}

@article{Kauffmann2003,
  author = {G. Kauffmann and others},
  title = {The Effect of Metallicity on the Initial Mass Function and Star Formation in Galaxies},
  journal = {Monthly Notices of the Royal Astronomical Society},
  volume = {346},
  year = {2003},
  pages = {1055-1072},
}

@ARTICLE{Weisz2011,
       author = {{Weisz}, Daniel R. and {Dolphin}, Andrew E. and {Skillman}, Evan D. and {Holtzman}, Jon and {Gilbert}, Karoline M. and {Dalcanton}, Julianne J. and {Williams}, Benjamin F.},
        title = "{The Star Formation Histories of Local Group Dwarf Galaxies. I. Hubble Space Telescope/Wide Field Planetary Camera 2 Observations}",
      journal = {The Astrophysical Journal},
         year = {2014},
        month = {July},
       volume = {789},
       number = {2},
          eid = {147},
        pages = {147},
          doi = {10.1088/0004-637X/789/2/147},
archivePrefix = {arXiv},
       eprint = {1404.7144},
 primaryClass = {astro-ph.GA},
       adsurl = {https://ui.adsabs.harvard.edu/abs/2014ApJ...789..147W}
}

@ARTICLE{2005astro.ph..6430D,
       author = {{Dolphin}, Andrew E. and {Weisz}, Daniel R. and {Skillman}, Evan D. and {Holtzman}, Jon A.},
        title = "{Star Formation Histories of Local Group Dwarf Galaxies}",
      journal = {arXiv e-prints},
         year = {2005},
        month = {June},
          eid = {astro-ph/0506430},
        pages = {astro-ph/0506430},
          doi = {10.48550/arXiv.astro-ph/0506430},
archivePrefix = {arXiv},
       eprint = {astro-ph/0506430},
 primaryClass = {astro-ph},
       adsurl = {https://ui.adsabs.harvard.edu/abs/2005astro.ph..6430D}
}

@article{Geha_2024,
   title={The SAGA Survey. IV. The Star Formation Properties of 101 Satellite Systems around Milky Way–mass Galaxies},
   volume={976},
   ISSN={1538-4357},
   url={http://dx.doi.org/10.3847/1538-4357/ad61e7},
   DOI={10.3847/1538-4357/ad61e7},
   number={1},
   journal={The Astrophysical Journal},
   publisher={American Astronomical Society},
   author={Geha, Marla and Mao, Yao-Yuan and Wechsler, Risa H. and Asali, Yasmeen and Kado-Fong, Erin and Kallivayalil, Nitya and Nadler, Ethan O. and Tollerud, Erik J. and Weiner, Benjamin and de los Reyes, Mithi A. C. and Wang, Yunchong and Wu, John F.},
   year={2024},
   month={November}, 
   pages={118}
}

@article{Dessart12,
	adsurl = {http://adsabs.harvard.edu/abs/2012arXiv1208.1214D},
	archiveprefix = {arXiv},
	author = {{Dessart}, L. and {Hillier}, D.~J. and {Waldman}, R. and {Livne}, E. and {Blondin}, S.},
	eprint = {1208.1214},
	journal = {ArXiv e-prints},
	month = aug,
	primaryclass = {astro-ph.SR},
	title = {{Super-luminous supernovae: 56Ni power versus magnetar radiation}},
	year = 2012}

@inproceedings{Quimby12,
	adsurl = {http://adsabs.harvard.edu/abs/2012IAUS..279...22Q},
	author = {{Quimby}, R.~M.},
	booktitle = {IAU Symposium},
	doi = {10.1017/S174392131201263X},
	month = sep,
	pages = {22-28},
	series = {IAU Symposium},
	title = {{Superluminous Supernovae}},
	volume = 279,
	year = 2012}

@article{Kasen10,
	adsurl = {http://adsabs.harvard.edu/abs/2010ApJ...717..245K},
	archiveprefix = {arXiv},
	author = {{Kasen}, D. and {Bildsten}, L.},
	doi = {10.1088/0004-637X/717/1/245},
	eprint = {0911.0680},
	journal = {ApJ},
	month = jul,
	pages = {245-249},
	primaryclass = {astro-ph.HE},
	title = {{Supernova Light Curves Powered by Young Magnetars}},
	volume = 717,
	year = 2010}

@article{Metzger_2015,
   title={The diversity of transients from magnetar birth in core collapse supernovae},
   volume={454},
   ISSN={1365-2966},
   url={http://dx.doi.org/10.1093/mnras/stv2224},
   DOI={10.1093/mnras/stv2224},
   number={3},
   journal={Monthly Notices of the Royal Astronomical Society},
   publisher={Oxford University Press (OUP)},
   author={Metzger, Brian D. and Margalit, Ben and Kasen, Daniel and Quataert, Eliot},
   year={2015},
   month=oct, pages={3311–3316} }

@article{Kashiyama_2016,
   title={MULTI-MESSENGER TESTS FOR FAST-SPINNING NEWBORN PULSARS EMBEDDED IN STRIPPED-ENVELOPE SUPERNOVAE},
   volume={818},
   ISSN={1538-4357},
   url={http://dx.doi.org/10.3847/0004-637X/818/1/94},
   DOI={10.3847/0004-637x/818/1/94},
   number={1},
   journal={The Astrophysical Journal},
   publisher={American Astronomical Society},
   author={Kashiyama, Kazumi and Murase, Kohta and Bartos, Imre and Kiuchi, Kenta and Margutti, Raffaella},
   year={2016},
   month=feb, pages={94} }

@article{KPO13,
	adsurl = {http://adsabs.harvard.edu/abs/2013MNRAS.432.3228K},
	archiveprefix = {arXiv},
	author = {{Kotera}, K. and {Phinney}, E.~S. and {Olinto}, A.~V.},
	doi = {10.1093/mnras/stt680},
	eprint = {1304.5326},
	journal = {MNRAS},
	month = jul,
	pages = {3228-3236},
	primaryclass = {astro-ph.HE},
	title = {{Signatures of pulsars in the light curves of newly formed supernova remnants}},
	volume = 432,
	year = 2013}

@article{Blasi00,
	adsurl = {http://adsabs.harvard.edu/abs/2000ApJ...533L.123B},
	author = {{Blasi}, P. and {Epstein}, R.~I. and {Olinto}, A.~V.},
	doi = {10.1086/312626},
	eprint = {arXiv:astro-ph/9912240},
	journal = {ApJ Lett.},
	month = apr,
	pages = {L123-L126},
	title = {{Ultra-High-Energy Cosmic Rays from Young Neutron Star Winds}},
	volume = 533,
	year = 2000}

@article{Bednarek97,
	adsurl = {http://adsabs.harvard.edu/abs/1997PhRvL..79.2616B},
	author = {{Bednarek}, W. and {Protheroe}, R.~J.},
	doi = {10.1103/PhysRevLett.79.2616},
	eprint = {arXiv:astro-ph/9704186},
	journal = {Physical Review Letters},
	month = oct,
	pages = {2616-2619},
	title = {{Gamma Rays and Neutrinos from the Crab Nebula Produced by Pulsar Accelerated Nuclei}},
	volume = 79,
	year = 1997}

@article{K11,
	adsurl = {http://adsabs.harvard.edu/abs/2011PhRvD..84b3002K},
	archiveprefix = {arXiv},
	author = {{Kotera}, K.},
	doi = {10.1103/PhysRevD.84.023002},
	eprint = {1106.3060},
	journal = {Phys. Rev. D},
	month = jul,
	number = 2,
	pages = {023002-+},
	primaryclass = {astro-ph.HE},
	title = {{Ultrahigh energy cosmic ray acceleration in newly born magnetars and their associated gravitational wave signatures}},
	volume = 84,
	year = 2011}

@ARTICLE{Fang15,
       author = {{Fang}, Ke},
        title = "{High-energy neutrino signatures of newborn pulsars in the local universe}",
      journal = {JCAP},
         year = 2015,
        month = jun,
       volume = {2015},
       number = {6},
        pages = {004-004},
          doi = {10.1088/1475-7516/2015/06/004},
archivePrefix = {arXiv},
       eprint = {1411.2174},
 primaryClass = {astro-ph.HE},
       adsurl = {https://ui.adsabs.harvard.edu/abs/2015JCAP...06..004F}
}

@ARTICLE{Fang14,
       author = {{Fang}, Ke and {Kotera}, Kumiko and {Murase}, Kohta and {Olinto}, Angela V.},
        title = "{Testing the newborn pulsar origin of ultrahigh energy cosmic rays with EeV neutrinos}",
      journal = {Phys. Rev. D},
         year = 2014,
        month = nov,
       volume = {90},
       number = {10},
          eid = {103005},
        pages = {103005},
          doi = {10.1103/PhysRevD.90.103005},
archivePrefix = {arXiv},
       eprint = {1311.2044},
 primaryClass = {astro-ph.HE},
       adsurl = {https://ui.adsabs.harvard.edu/abs/2014PhRvD..90j3005F}
}

@article{Beniamini_2019,
   title={Formation rates and evolution histories of magnetars},
   volume={487},
   ISSN={1365-2966},
   url={http://dx.doi.org/10.1093/mnras/stz1391},
   DOI={10.1093/mnras/stz1391},
   number={1},
   journal={Monthly Notices of the Royal Astronomical Society},
   publisher={Oxford University Press (OUP)},
   author={Beniamini, Paz and Hotokezaka, Kenta and van der Horst, Alexander and Kouveliotou, Chryssa},
   year={2019},
   month=may, pages={1426–1438} }

@techreport{Gen2_TDR,
    collaboration = "IceCube-Gen2",
    title = "{IceCube-Gen2 Technical Design Report}",
    url = {https://icecube-gen2.wisc.edu/science/publications/tdr/},
    journal="",
    year = {2024},
    number=""
}

@article{Zeolla:ARENA2024,
  author = "Zeolla, Andrew  and others",
  title = "{Sensitivity of BEACON to Ultrahigh Energy Neutrinos}",
  journal = "PoS",
  year = 2024,
  volume = "ARENA2024",
  pages = "039"
}

@inproceedings{Kotera:2024iyk,
    author = "Kotera, Kumiko",
    collaboration = "GRAND",
    title = "{GRAND: status and perspectives}",
    booktitle = "{10th International Workshop on Acoustic and Radio EeV Neutrino Detection Activities}",
    eprint = "2408.16316",
    archivePrefix = "arXiv",
    primaryClass = "astro-ph.HE",
    month = "8",
    year = "2024"
}

@article{Feldman_1998,
   title={Unified approach to the classical statistical analysis of small signals},
   volume={57},
   ISSN={1089-4918},
   url={http://dx.doi.org/10.1103/PhysRevD.57.3873},
   DOI={10.1103/physrevd.57.3873},
   number={7},
   journal={Physical Review D},
   publisher={American Physical Society (APS)},
   author={Feldman, Gary J. and Cousins, Robert D.},
   year={1998},
   month=apr, pages={3873–3889} }

@misc{bouma2023directionreconstructioniniceradio,
      title={Direction reconstruction for the in-ice radio array of IceCube-Gen2}, 
      author={Sjoerd Bouma and Anna Nelles},
      year={2023},
      eprint={2307.13971},
      archivePrefix={arXiv},
      primaryClass={astro-ph.HE},
      url={https://arxiv.org/abs/2307.13971} 
}

@ARTICLE{Piran2005,
       author = {{Piran}, T.},
        title = "{The beaming factor and other open issues in GRB Jets}",
      journal = {Nuovo Cimento C Geophysics Space Physics C},
         year = 2005,
        month = may,
       volume = {28},
       number = {3},
        pages = {373},
          doi = {10.1393/ncc/i2005-10063-y},
archivePrefix = {arXiv},
       eprint = {astro-ph/0502473},
 primaryClass = {astro-ph},
       adsurl = {https://ui.adsabs.harvard.edu/abs/2005NCimC..28..373P}
}

@article{Murase_2010,
   title={Neutrino background flux from sources of ultrahigh-energy cosmic-ray nuclei},
   volume={81},
   ISSN={1550-2368},
   url={http://dx.doi.org/10.1103/PhysRevD.81.123001},
   DOI={10.1103/physrevd.81.123001},
   number={12},
   journal={Physical Review D},
   publisher={American Physical Society (APS)},
   author={Murase, Kohta and Beacom, John F.},
   year={2010},
   month=jun }

@article{Ehlert_2023,
   title={Curious case of the maximum rigidity distribution of cosmic-ray accelerators},
   volume={107},
   ISSN={2470-0029},
   url={http://dx.doi.org/10.1103/PhysRevD.107.103045},
   DOI={10.1103/physrevd.107.103045},
   number={10},
   journal={Physical Review D},
   publisher={American Physical Society (APS)},
   author={Ehlert, D. and Oikonomou, F. and Unger, M.},
   year={2023},
   month=may }

@article{Fang_2016,
   title={A NEW METHOD FOR FINDING POINT SOURCES IN HIGH-ENERGY NEUTRINO DATA},
   volume={826},
   ISSN={1538-4357},
   url={http://dx.doi.org/10.3847/0004-637X/826/2/102},
   DOI={10.3847/0004-637x/826/2/102},
   number={2},
   journal={The Astrophysical Journal},
   publisher={American Astronomical Society},
   author={Fang, Ke and Miller, M. Coleman},
   year={2016},
   month=jul, pages={102} }

@article{valera2023,
    author = "Valera, Victor Branco and Bustamante, Mauricio and Glaser, Christian",
    title = "{Near-future discovery of the diffuse flux of ultrahigh-energy cosmic neutrinos}",
    eprint = "2210.03756",
    archivePrefix = "arXiv",
    primaryClass = "astro-ph.HE",
    doi = "10.1103/PhysRevD.107.043019",
    journal = "Phys. Rev. D",
    volume = "107",
    number = "4",
    pages = "043019",
    year = "2023"
}

@misc{lundquist2024,
      title={Combined Fit of Spectrum and Composition for FR0 Radio Galaxy Emitted Ultra-High-Energy Cosmic Rays with Resulting Secondary Photons and Neutrinos}, 
      author={Jon Paul Lundquist and Serguei Vorobiov and Lukas Merten and Anita Reimer and Margot Boughelilba and Paolo Da Vela and Fabrizio Tavecchio and Giacomo Bonnoli and Chiara Righi},
      year={2024},
      eprint={2407.06961},
      archivePrefix={arXiv},
      primaryClass={astro-ph.HE},
      url={https://arxiv.org/abs/2407.06961}, 
}

@article{Heinze_2019,
   title={A New View on Auger Data and Cosmogenic Neutrinos in Light of Different Nuclear Disintegration and Air-shower Models},
   volume={873},
   ISSN={1538-4357},
   url={http://dx.doi.org/10.3847/1538-4357/ab05ce},
   DOI={10.3847/1538-4357/ab05ce},
   number={1},
   journal={The Astrophysical Journal},
   publisher={American Astronomical Society},
   author={Heinze, Jonas and Fedynitch, Anatoli and Boncioli, Denise and Winter, Walter},
   year={2019},
   month=mar, pages={88} }

@article{Aab_2017,
   title={Combined fit of spectrum and composition data as measured by the Pierre Auger Observatory},
   volume={2017},
   ISSN={1475-7516},
   url={http://dx.doi.org/10.1088/1475-7516/2017/04/038},
   DOI={10.1088/1475-7516/2017/04/038},
   number={04},
   journal={Journal of Cosmology and Astroparticle Physics},
   publisher={IOP Publishing},
   author={Aab, A. and Abreu, P. and Aglietta, M. and others},
   year={2017},
   month=apr, pages={038–038} }

@misc{gonzalez2022,
      title={Searches for ultra-high energy photons and neutrinos with the Pierre Auger Observatory}, 
      author={Nicolás Martín González},
      year={2022},
      eprint={2212.07139},
      archivePrefix={arXiv},
      primaryClass={astro-ph.HE},
      url={https://arxiv.org/abs/2212.07139}, 
}

@article{IC_Science_2022,
   title={Evidence for neutrino emission from the nearby active galaxy NGC 1068},
   volume={378},
   ISSN={1095-9203},
   url={http://dx.doi.org/10.1126/science.abg3395},
   DOI={10.1126/science.abg3395},
   number={6619},
   journal={Science},
   publisher={American Association for the Advancement of Science (AAAS)},
   author={Abbasi, R. and Ackermann, M. and Adams, J. and others},
   year={2022},
   month=nov, pages={538–543} }

@article{IceCube_Science_2023,
   title={Observation of high-energy neutrinos from the Galactic plane},
   volume={380},
   ISSN={1095-9203},
   url={http://dx.doi.org/10.1126/science.adc9818},
   DOI={10.1126/science.adc9818},
   number={6652},
   journal={Science},
   publisher={American Association for the Advancement of Science (AAAS)},
   author={Abbasi, R. and Ackermann, M. and Adams, J. and others},
   year={2023},
   month=jun, pages={1338–1343} }

@ARTICLE{2022NatRP...4..697G,
       author = {{Gu{\'e}pin}, Claire and {Kotera}, Kumiko and {Oikonomou}, Foteini},
        title = "{High-energy neutrino transients and the future of multi-messenger astronomy}",
      journal = {Nature Reviews Physics},
         year = 2022,
        month = nov,
       volume = {4},
       number = {11},
        pages = {697-712},
          doi = {10.1038/s42254-022-00504-9},
archivePrefix = {arXiv},
       eprint = {2207.12205},
 primaryClass = {astro-ph.HE},
       adsurl = {https://ui.adsabs.harvard.edu/abs/2022NatRP...4..697G}
}

@article{Alves_2019,
   title={Cosmogenic photon and neutrino fluxes in the Auger era},
   volume={2019},
   ISSN={1475-7516},
   url={http://dx.doi.org/10.1088/1475-7516/2019/01/002},
   DOI={10.1088/1475-7516/2019/01/002},
   number={01},
   journal={Journal of Cosmology and Astroparticle Physics},
   publisher={IOP Publishing},
   author={Alves Batista, R. and de Almeida, Rogerio M. and Lago, Bruno and Kotera, Kumiko},
   year={2019},
   month=jan, pages={002–002} }

@article{Decoene_2023,
   title={Radio wavefront of very inclined extensive air-showers: A simulation study for extended and sparse radio arrays},
   volume={145},
   ISSN={0927-6505},
   url={http://dx.doi.org/10.1016/j.astropartphys.2022.102779},
   DOI={10.1016/j.astropartphys.2022.102779},
   journal={Astroparticle Physics},
   publisher={Elsevier BV},
   author={Decoene, Valentin and Martineau-Huynh, Olivier and Tueros, Matias},
   year={2023},
   month=mar, pages={102779} }

@article{decoene2023reviewneutrinoexperimentssearching,
    author = "Decoene, Valentin",
    title = "{Review of Neutrino Experiments Searching for Astrophysical Neutrinos}",
    eprint = "2309.17139",
    archivePrefix = "arXiv",
    primaryClass = "astro-ph.HE",
    doi = "10.22323/1.444.0026",
    journal = "PoS",
    volume = "ICRC2023",
    pages = "026",
    year = "2023"
}

@ARTICLE{2017Sci...357.1266P,
       author = {{Pierre Auger Collaboration} and others},
        title = "{Observation of a large-scale anisotropy in the arrival directions of cosmic rays above 8 {\texttimes} {}10$^{18}$ eV}",
      journal = {Science},
         year = 2017,
        month = sep,
       volume = {357},
       number = {6357},
        pages = {1266-1270},
          doi = {10.1126/science.aan4338},
archivePrefix = {arXiv},
       eprint = {1709.07321},
 primaryClass = {astro-ph.HE},
       adsurl = {https://ui.adsabs.harvard.edu/abs/2017Sci...357.1266P}
}

@article{Aartsen_2018,
   title={Differential limit on the extremely-high-energy cosmic neutrino flux in the presence of astrophysical background from nine years of IceCube data},
   volume={98},
   ISSN={2470-0029},
   url={http://dx.doi.org/10.1103/PhysRevD.98.062003},
   DOI={10.1103/physrevd.98.062003},
   number={6},
   journal={Physical Review D},
   publisher={American Physical Society (APS)},
   author={Aartsen, M. G. and Ackermann, M. and Adams, J. and others},
   year={2018},
   month=sep }

@article{IceCube21,
   title={IceCube high-energy starting event sample: Description and flux characterization with 7.5 years of data},
   volume={104},
   ISSN={2470-0029},
   url={http://dx.doi.org/10.1103/PhysRevD.104.022002},
   DOI={10.1103/physrevd.104.022002},
   number={2},
   journal={Physical Review D},
   publisher={American Physical Society (APS)},
   author={Abbasi, R. and Ackermann, M. and Adams, J. and Aguilar, J. A. and Ahlers, M. and Ahrens, M. and Alispach, C. and Alves, A. A. and Amin, N. M. and Andeen, K. and et al.},
   year={2021},
   month={Jul}
}

@article{IceCube:2021uhz,
    author = "Abbasi, R. and others",
    collaboration = "IceCube",
    title = "{Improved Characterization of the Astrophysical Muon-Neutrino Flux with 9.5 Years of IceCube Data}",
    eprint = "2111.10299",
    archivePrefix = "arXiv",
    primaryClass = "astro-ph.HE",
   doi = "10.3847/1538-4357/ac4d29",
    journal = "Astrophys. J.",
    volume = "928",
    number = "1",
    pages = "50",
    year = "2022"
}

@ARTICLE{2020PhRvL.124e1103A,
       author = {{Aartsen}, M.~G. and {Ackermann}, M. and {Adams}, J. and others},
        title = "{Time-Integrated Neutrino Source Searches with 10 Years of IceCube Data}",
      journal = {Phys. Rev. Letters},
         year = 2020,
        month = feb,
       volume = {124},
       number = {5},
          eid = {051103},
        pages = {051103},
          doi = {10.1103/PhysRevLett.124.051103},
archivePrefix = {arXiv},
       eprint = {1910.08488},
 primaryClass = {astro-ph.HE},
       adsurl = {https://ui.adsabs.harvard.edu/abs/2020PhRvL.124e1103A}
}

@article{Barwick:2021wU,
  author = "Barwick, Steve  and  for the ARIANNA collaboration",
  title = "{Capabilities of ARIANNA: Neutrino Pointing Resolution and Implications for Future Ultra-high Energy Neutrino Astronomy}",
  doi = "10.22323/1.395.1151",
  journal = "PoS",
  year = 2021,
  volume = "ICRC2021",
  pages = "1151"
}

@article{Aab_2019,
	doi = {10.1088/1475-7516/2019/10/022},
	url = {https://doi.org/10.1088/1475-7516/2019/10/022},
	year = 2019,
	month = {oct},
	publisher = {{IOP} Publishing},
	volume = {2019},
	number = {10},
	pages = {022--022},
	author = {A. Aab and P. Abreu and M. Aglietta and others},
	title = {Probing the origin of ultra-high-energy cosmic rays with neutrinos in the {EeV} energy range using the Pierre Auger Observatory},
	journal = {Journal of Cosmology and Astroparticle Physics}
}

@article{2020arXiv200804323T,
    author = "Aartsen, M. G. and others",
    collaboration = "IceCube-Gen2",
    title = "{IceCube-Gen2: the window to the extreme Universe}",
    eprint = "2008.04323",
    archivePrefix = "arXiv",
    primaryClass = "astro-ph.HE",
    doi = "10.1088/1361-6471/abbd48",
    journal = "J. Phys. G",
    volume = "48",
    number = "6",
    pages = "060501",
    year = "2021"
}

@ARTICLE{2021JInst..16P8035A,
       author = {{The PUEO collaboration}},
        title = "{The Payload for Ultrahigh Energy Observations (PUEO): a white paper}",
      journal = {Journal of Instrumentation},
         year = 2021,
        month = aug,
       volume = {16},
       number = {8},
          eid = {P08035},
        pages = {P08035},
          doi = {10.1088/1748-0221/16/08/P08035},
archivePrefix = {arXiv},
       eprint = {2010.02892},
 primaryClass = {astro-ph.IM},
       adsurl = {https://ui.adsabs.harvard.edu/abs/2021JInst..16P8035A}
}

@INPROCEEDINGS{2021cosp...43E1367O,
       author = {{Olinto}, Angela V.},
        title = "{The POEMMA (Probe Of Extreme Multi-Messenger Astrophysics) mission}",
    booktitle = {43rd COSPAR Scientific Assembly. Held 28 January - 4 February},
         year = 2021,
       volume = {43},
        month = jan,
          eid = {1367},
        pages = {1367},
archivePrefix = {arXiv},
       eprint = {2012.07945},
 primaryClass = {astro-ph.IM},
       adsurl = {https://ui.adsabs.harvard.edu/abs/2021cosp...43E1367O}
}

@ARTICLE{2021arXiv210802751W,
       author = {{Wang}, Andrew and {Lin}, Chaoxian and {Otte}, Nepomuk and {Doro}, Michele and {Gazda}, Eliza and {Taboada}, Ignacio and {Brown}, Anthony and {Bagheri}, Mahdi},
        title = "{Trinity's Sensitivity to Isotropic and Point-Source Neutrinos}",
         year = 2021,
          doi = "10.22323/1.395.1234",
    journal = "PoS",
    volume = "ICRC2021",
    pages = "1234",
          eid = {arXiv:2108.02751},
        pages = {arXiv:2108.02751},
archivePrefix = {arXiv},
       eprint = {2108.02751},
 primaryClass = {astro-ph.IM},
       adsurl = {https://ui.adsabs.harvard.edu/abs/2021arXiv210802751W}
}

@ARTICLE{2020arXiv200206475R,
       author = {{Romero-Wolf}, Andres and {Alvarez-Mu{\~n}iz}, Jaime and {Carvalho}, Washington R., Jr. and others},
        title = "{An Andean Deep-Valley Detector for High-Energy Tau Neutrinos}",
         year = 2020,
        month = feb,
        journal = "{Latin American Strategy Forum for Research Infrastructure}",
          eid = {arXiv:2002.06475},
        pages = {arXiv:2002.06475},
archivePrefix = {arXiv},
       eprint = {2002.06475},
 primaryClass = {astro-ph.IM},
       adsurl = {https://ui.adsabs.harvard.edu/abs/2020arXiv200206475R}
}

@ARTICLE{2019ARNPS..69..477M,
       author = {{Murase}, Kohta and {Bartos}, Imre},
        title = "{High-Energy Multimessenger Transient Astrophysics}",
      journal = {Annual Review of Nuclear and Particle Science},
         year = 2019,
        month = oct,
       volume = {69},
        pages = {477-506},
          doi = {10.1146/annurev-nucl-101918-023510},
archivePrefix = {arXiv},
       eprint = {1907.12506},
 primaryClass = {astro-ph.HE},
       adsurl = {https://ui.adsabs.harvard.edu/abs/2019ARNPS..69..477M}
}

@article{Abreu13_18,
	Adsurl = {http://adsabs.harvard.edu/abs/2013ApJ...762L..13P},
	Archiveprefix = {arXiv},
	Author = {{Pierre Auger Collaboration} and {Abreu}, P. and {Aglietta}, M. and {Ahlers}, M. and {Ahn}, E.~J. and {Albuquerque}, I.~F.~M. and {Allard}, D. and {Allekotte}, I. and {Allen}, J. and {Allison}, P. and et al.},
	Doi = {10.1088/2041-8205/762/1/L13},
	Eid = {L13},
	Eprint = {1212.3083},
	Journal = {ApJ Lett.},
	Month = jan,
	Pages = {L13-21},
	Primaryclass = {astro-ph.HE},
	Title = {{Constraints on the Origin of Cosmic Rays above 10$^{18}$ eV from Large-scale Anisotropy Searches in Data of the Pierre Auger Observatory}},
	Volume = 762,
	Year = 2013}

@article{Palladino_2020,
   title={Can astrophysical neutrinos trace the origin of the detected ultra-high energy cosmic rays?},
   volume={494},
   ISSN={1365-2966},
   url={http://dx.doi.org/10.1093/mnras/staa1003},
   DOI={10.1093/mnras/staa1003},
   number={3},
   journal={Monthly Notices of the Royal Astronomical Society},
   publisher={Oxford University Press (OUP)},
   author={Palladino, Andrea and van Vliet, Arjen and Winter, Walter and Franckowiak, Anna},
   year={2020},
   month={Apr},
   pages={4255–4265}
}

@ARTICLE{2019FrASS...6...23B,
       author = {{Alves Batista}, Rafael and {Biteau}, Jonathan and {Bustamante}, Mauricio and {Dolag}, Klaus and {Engel}, Ralph and {Fang}, Ke and {Kampert}, Karl-Heinz and {Kostunin}, Dmitriy and {Mostafa}, Miguel and {Murase}, Kohta and {Oikonomou}, Foteini and {Olinto}, Angela V. and {Panasyuk}, Mikhail I. and {Sigl}, Guenter and {Taylor}, Andrew M. and {Unger}, Michael},
        title = "{Open questions in cosmic-ray research at ultrahigh energies}",
      journal = {Frontiers in Astronomy and Space Sciences},
         year = 2019,
        month = jun,
       volume = {6},
          eid = {23},
        pages = {23},
          doi = {10.3389/fspas.2019.00023},
archivePrefix = {arXiv},
       eprint = {1903.06714},
 primaryClass = {astro-ph.HE},
       adsurl = {https://ui.adsabs.harvard.edu/abs/2019FrASS...6...23B}
}

@ARTICLE{2016ApJ...832L..17F,
       author = {{Fang}, Ke and {Kotera}, Kumiko},
        title = "{The Highest-energy Cosmic Rays Cannot Be Dominantly Protons from Steady Sources}",
      journal = {Astrophysical Journal Letters},
         year = 2016,
        month = nov,
       volume = {832},
       number = {1},
          eid = {L17},
        pages = {L17},
          doi = {10.3847/2041-8205/832/1/L17},
archivePrefix = {arXiv},
       eprint = {1610.08055},
 primaryClass = {astro-ph.HE},
       adsurl = {https://ui.adsabs.harvard.edu/abs/2016ApJ...832L..17F}
}

@ARTICLE{2016PhRvD..94j3006M,
       author = {{Murase}, Kohta and {Waxman}, Eli},
        title = "{Constraining high-energy cosmic neutrino sources: Implications and prospects}",
      journal = {Physical Review D},
         year = 2016,
        month = nov,
       volume = {94},
       number = {10},
          eid = {103006},
        pages = {103006},
          doi = {10.1103/PhysRevD.94.103006},
archivePrefix = {arXiv},
       eprint = {1607.01601},
 primaryClass = {astro-ph.HE},
       adsurl = {https://ui.adsabs.harvard.edu/abs/2016PhRvD..94j3006M}
}

@ARTICLE{Yoshida22,
       author = {{Yoshida}, Shigeru and {Murase}, Kohta and {Tanaka}, Masaomi and {Shimizu}, Nobuhiro and {Ishihara}, Aya},
        title = "{Identifying High-energy Neutrino Transients by Neutrino Multiplet-triggered Follow-ups}",
      journal = {The Astrophysical Journal},
         year = 2022,
        month = oct,
       volume = {937},
       number = {2},
          eid = {108},
        pages = {108},
          doi = {10.3847/1538-4357/ac8dfd},
archivePrefix = {arXiv},
       eprint = {2206.13719},
 primaryClass = {astro-ph.HE},
       adsurl = {https://ui.adsabs.harvard.edu/abs/2022ApJ...937..108Y}
}

@ARTICLE{2017ApJ...850L..35A,
       author = {{Albert}, A. and {Andr{\'e}}, M. and {Anghinolfi}, M. and others},
        title = "{Search for High-energy Neutrinos from Binary Neutron Star Merger GW170817 with ANTARES, IceCube, and the Pierre Auger Observatory}",
      journal = {Astrophysical Journal Letters},
         year = 2017,
        month = dec,
       volume = {850},
       number = {2},
          eid = {L35},
        pages = {L35},
          doi = {10.3847/2041-8213/aa9aed},
archivePrefix = {arXiv},
       eprint = {1710.05839},
 primaryClass = {astro-ph.HE},
       adsurl = {https://ui.adsabs.harvard.edu/abs/2017ApJ...850L..35A}
}

@article{2013PhRvL.111b1103A,
	Adsurl = {http://adsabs.harvard.edu/abs/2013PhRvL.111b1103A},
	Archiveprefix = {arXiv},
	Author = {{Aartsen}, M.~G. and {Abbasi}, R. and {Abdou}, Y. and {Ackermann}, M. and {Adams}, J. and {Aguilar}, J.~A. and {Ahlers}, M. and {Altmann}, D. and {Auffenberg}, J. and {Bai}, X. and et al.},
	Doi = {10.1103/PhysRevLett.111.021103},
	Eid = {021103},
	Eprint = {1304.5356},
	Journal = {Physical Review Letters},
	Month = jul,
	Number = 2,
	Pages = {021103},
	Primaryclass = {astro-ph.HE},
	Title = {{First Observation of PeV-Energy Neutrinos with IceCube}},
	Volume = 111,
	Year = 2013}

@ARTICLE{Stein20,
       author = {{Stein}, Robert and {van Velzen}, Sjoert and {Kowalski}, Marek and {Franckowiak}, Anna and {Gezari}, Suvi and others},
    title = "{A tidal disruption event coincident with a high-energy neutrino}",
    eprint = "2005.05340",
    archivePrefix = "arXiv",
    primaryClass = "astro-ph.HE",
    doi = "10.1038/s41550-020-01295-8",
    journal = "Nature Astron.",
    volume = "5",
    number = "5",
    pages = "510--518",
    year = "2021"
}

@phdthesis{DecoenePhD20,
  TITLE = {{Sources and detection of high energy cosmic events~}},
  AUTHOR = {Decoene, Valentin},
  URL = {https://hal.archives-ouvertes.fr/tel-02991529},
  SCHOOL = {{Institut d'Astrophysique de Paris, Sorbonne Universit{\'e}}},
  YEAR = {2020},
  MONTH = Sep,
  TYPE = {Theses},
  HAL_ID = {tel-02991529},
  HAL_VERSION = {v1},
}

@article{Guepin17,
	adsurl = {http://adsabs.harvard.edu/abs/2017A%26A...603A..76G},
	archiveprefix = {arXiv},
	author = {{Gu{\'e}pin}, C. and {Kotera}, K.},
	doi = {10.1051/0004-6361/201630326},
	eid = {A76},
	eprint = {1701.07038},
	journal = {Astronomy \& Astrophysics},
	month = jul,
	pages = {A76},
	primaryclass = {astro-ph.HE},
	title = {{Can we observe neutrino flares in coincidence with explosive transients?}},
	volume = 603,
	year = 2017}

@article{Kotera09,
	adsurl = {http://adsabs.harvard.edu/abs/2009ApJ...707..370K},
	archiveprefix = {arXiv},
	author = {{Kotera}, K. and {Allard}, D. and {Murase}, K. and {Aoi}, J. and {Dubois}, Y. and {Pierog}, T. and {Nagataki}, S.},
	doi = {10.1088/0004-637X/707/1/370},
	eprint = {0907.2433},
	journal = {ApJ},
	month = dec,
	pages = {370-386},
	title = {{Propagation of Ultrahigh Energy Nuclei in Clusters of Galaxies: Resulting Composition and Secondary Emissions}},
	volume = 707,
	year = 2009}

@article{Rodrigues21,
   title={Active Galactic Nuclei Jets as the Origin of Ultrahigh-Energy Cosmic Rays and Perspectives for the Detection of Astrophysical Source Neutrinos at EeV Energies},
   volume={126},
   ISSN={1079-7114},
   url={http://dx.doi.org/10.1103/PhysRevLett.126.191101},
   DOI={10.1103/physrevlett.126.191101},
   number={19},
   journal={Physical Review Letters},
   publisher={American Physical Society (APS)},
   author={Rodrigues, Xavier and Heinze, Jonas and Palladino, Andrea and van Vliet, Arjen and Winter, Walter},
   year={2021},
   month={May}
}

@article{Decoene:2019eux,
    author = "Decoene, Valentin and Gu\'epin, C. and Fang, K. and Kotera, K. and Metzger, B. D.",
    title = "{High-energy neutrinos from fallback accretion of binary neutron star merger remnants}",
    eprint = "1910.06578",
    archivePrefix = "arXiv",
    primaryClass = "astro-ph.HE",
    doi = "10.1088/1475-7516/2020/04/045",
    journal = "JCAP",
    volume = "04",
    pages = "045",
    year = "2020"
}

@article{PhysRevD.90.103005,
	Author = {Fang, Ke and Kotera, Kumiko and Murase, Kohta and Olinto, Angela V.},
	Doi = {10.1103/PhysRevD.90.103005},
	Issue = {10},
	Journal = {Phys. Rev. D},
	Month = {Nov},
	Numpages = {7},
	Pages = {103005},
	Publisher = {American Physical Society},
	Title = {Testing the newborn pulsar origin of ultrahigh energy cosmic rays with EeV neutrinos},
	Url = {https://link.aps.org/doi/10.1103/PhysRevD.90.103005},
	Volume = {90},
	Year = {2014}}

@article{Fang:2016hop,
	Archiveprefix = {arXiv},
	Author = {Fang, Ke and Kotera, Kumiko and Miller, M. Coleman and others},
	Doi = {10.1088/1475-7516/2016/12/017},
	Eprint = {1609.08027},
	Journal = {JCAP},
	Number = {12},
	Pages = {017},
	Primaryclass = {astro-ph.HE},
	Title = {{Identifying Ultrahigh-Energy Cosmic-Ray Accelerators with Future Ultrahigh-Energy Neutrino Detectors}},
	Volume = {1612},
	Year = {2016}}

@article{Fang:2017tla,
    author = "Fang, Ke and Metzger, Brian D.",
    title = "{High-Energy Neutrinos from Millisecond Magnetars formed from the Merger of Binary Neutron Stars}",
    eprint = "1707.04263",
    archivePrefix = "arXiv",
    primaryClass = "astro-ph.HE",
    doi = "10.3847/1538-4357/aa8b6a",
    journal = "Astrophys. J.",
    volume = "849",
    number = "2",
    pages = "153",
    year = "2017"
}

@article{IceCube:2018dnn,
    author = "Aartsen, M. G. and others",
    collaboration = "IceCube, Fermi-LAT, MAGIC, AGILE, ASAS-SN, HAWC, H.E.S.S., INTEGRAL, Kanata, Kiso, Kapteyn, Liverpool Telescope, Subaru, Swift NuSTAR, VERITAS, VLA/17B-403",
    title = "{Multimessenger observations of a flaring blazar coincident with high-energy neutrino IceCube-170922A}",
    eprint = "1807.08816",
    archivePrefix = "arXiv",
    primaryClass = "astro-ph.HE",
    doi = "10.1126/science.aat1378",
    journal = "Science",
    volume = "361",
    number = "6398",
    pages = "eaat1378",
    year = "2018"
}

@article{IceCube:2018cha,
    author = "Aartsen, M. G. and others",
    collaboration = "IceCube",
    title = "{Neutrino emission from the direction of the blazar TXS 0506+056 prior to the IceCube-170922A alert}",
    eprint = "1807.08794",
    archivePrefix = "arXiv",
    primaryClass = "astro-ph.HE",
    doi = "10.1126/science.aat2890",
    journal = "Science",
    volume = "361",
    number = "6398",
    pages = "147--151",
    year = "2018"
}

@article{Arons03,
	adsurl = {http://adsabs.harvard.edu/abs/2003ApJ...589..871A},
	author = {{Arons}, J.},
	doi = {10.1086/374776},
	eprint = {astro-ph/0208444},
	journal = {The Astrophysical Journal},
	month = jun,
	pages = {871-892},
	title = {{Magnetars in the Metagalaxy: An Origin for Ultra-High-Energy Cosmic Rays in the Nearby Universe}},
	volume = 589,
	year = 2003}

@article{Fang12,
	adsurl = {http://adsabs.harvard.edu/abs/2012ApJ...750..118F},
	archiveprefix = {arXiv},
	author = {{Fang}, K. and {Kotera}, K. and {Olinto}, A.~V.},
	doi = {10.1088/0004-637X/750/2/118},
	eid = {118},
	eprint = {1201.5197},
	journal = {The Astrophysical Journal},
	month = may,
	pages = {118},
	primaryclass = {astro-ph.HE},
	title = {{Newly Born Pulsars as Sources of Ultrahigh Energy Cosmic Rays}},
	volume = 750,
	year = 2012}

@article{Kimura:2017kan,
    author = "Kimura, Shigeo S. and Murase, Kohta and M\'esz\'aros, Peter and Kiuchi, Kenta",
    title = "{High-Energy Neutrino Emission from Short Gamma-Ray Bursts: Prospects for Coincident Detection with Gravitational Waves}",
    eprint = "1708.07075",
    archivePrefix = "arXiv",
    primaryClass = "astro-ph.HE",
    doi = "10.3847/2041-8213/aa8d14",
    journal = "Astrophys. J. Lett.",
    volume = "848",
    number = "1",
    pages = "L4",
    year = "2017"
}

@article{Stettner:2019tok,
    author = "Stettner, J.",
    collaboration = "IceCube",
    title = "{Measurement of the Diffuse Astrophysical Muon-Neutrino Spectrum with Ten Years of IceCube Data}",
    eprint = "1908.09551",
    archivePrefix = "arXiv",
    primaryClass = "astro-ph.HE",
    reportNumber = "PoS-ICRC2019-1017",
    doi = "10.22323/1.358.1017",
    journal = "PoS",
    volume = "ICRC2019",
    pages = "1017",
    year = "2020"
}

@article{Murase:2009pg,
    author = "Murase, Kohta and Meszaros, Peter and Zhang, Bing",
    title = "{Probing the birth of fast rotating magnetars through high-energy neutrinos}",
    eprint = "0904.2509",
    archivePrefix = "arXiv",
    primaryClass = "astro-ph.HE",
    reportNumber = "YITP-09-30",
    doi = "10.1103/PhysRevD.79.103001",
    journal = "Phys. Rev. D",
    volume = "79",
    pages = "103001",
    year = "2009"
}

@article{Carpio:2020wzg,
    author = "Carpio, Jose Alonso and Murase, Kohta and Reno, Mary Hall and Sarcevic, Ina and Stasto, Anna",
    title = "{Charm contribution to ultrahigh-energy neutrinos from newborn magnetars}",
    eprint = "2007.07945",
    archivePrefix = "arXiv",
    primaryClass = "astro-ph.HE",
    doi = "10.1103/PhysRevD.102.103001",
    journal = "Phys. Rev. D",
    volume = "102",
    number = "10",
    pages = "103001",
    year = "2020"
}

@ARTICLE{WaxmanBahcall98,
       author = {{Waxman}, Eli and {Bahcall}, John},
        title = "{High energy neutrinos from astrophysical sources: An upper bound}",
      journal = {Physical Review D},
         year = 1998,
        month = dec,
       volume = {59},
       number = {2},
          eid = {023002},
        pages = {023002},
          doi = {10.1103/PhysRevD.59.023002},
archivePrefix = {arXiv},
       eprint = {hep-ph/9807282},
 primaryClass = {hep-ph},
       adsurl = {https://ui.adsabs.harvard.edu/abs/1998PhRvD..59b3002W}
}

@article{Murase:2007yt,
    author = "Murase, Kohta",
    title = "{High energy neutrino early afterglows gamma-ray bursts revisited}",
    eprint = "0707.1140",
    archivePrefix = "arXiv",
    primaryClass = "astro-ph",
    doi = "10.1103/PhysRevD.76.123001",
    journal = "Phys. Rev. D",
    volume = "76",
    pages = "123001",
    year = "2007"
}

@article{Anchordoqui:2018qom,
    author = "Anchordoqui, Luis A.",
    title = "{Ultra-High-Energy Cosmic Rays}",
    eprint = "1807.09645",
    archivePrefix = "arXiv",
    primaryClass = "astro-ph.HE",
    doi = "10.1016/j.physrep.2019.01.002",
    journal = "Phys. Rept.",
    volume = "801",
    pages = "1--93",
    year = "2019"
}

@article{Reusch:2021ztx,
    author = "Reusch, Simeon and others",
    title = "{The candidate tidal disruption event AT2019fdr coincident with a high-energy neutrino}",
    eprint = "2111.09390",
    archivePrefix = "arXiv",
    primaryClass = "astro-ph.HE",
    journal = "arXiv",
    month = "11",
    year = "2021"
}

@article{Fang:2017zjf,
    author = "Fang, Ke and Murase, Kohta",
    title = "{Linking High-Energy Cosmic Particles by Black Hole Jets Embedded in Large-Scale Structures}",
    eprint = "1704.00015",
    archivePrefix = "arXiv",
    primaryClass = "astro-ph.HE",
    doi = "10.1038/s41567-017-0025-4",
    journal = "Nature Phys.",
    volume = "14",
    number = "4",
    pages = "396--398",
    year = "2018"
}

@article{Mukhopadhyay:2024lwq,
    author = "Mukhopadhyay, Mainak and Kotera, Kumiko and Wissel, Stephanie and Murase, Kohta and Kimura, Shigeo S.",
    title = "{Ultrahigh-energy neutrino searches using next-generation gravitational wave detectors at radio neutrino detectors: GRAND, IceCube-Gen2 Radio, and RNO-G}",
    eprint = "2406.19440",
    archivePrefix = "arXiv",
    primaryClass = "astro-ph.HE",
    doi = "10.1103/PhysRevD.110.063004",
    journal = "Phys. Rev. D",
    volume = "110",
    number = "6",
    pages = "063004",
    year = "2024"
}

@article{Mukhopadhyay:2023niv,
    author = "Mukhopadhyay, Mainak and Kimura, Shigeo S. and Murase, Kohta",
    title = "{Gravitational wave triggered searches for high-energy neutrinos from binary neutron star mergers: Prospects for next generation detectors}",
    eprint = "2310.16875",
    archivePrefix = "arXiv",
    primaryClass = "astro-ph.HE",
    doi = "10.1103/PhysRevD.109.043053",
    journal = "Phys. Rev. D",
    volume = "109",
    number = "4",
    pages = "043053",
    year = "2024"
}

@article{Mukhopadhyay:2024ehs,
    author = "Mukhopadhyay, Mainak and Kimura, Shigeo S. and Metzger, Brian D.",
    title = "{High-energy neutrino signatures from pulsar remnants of binary neutron-star mergers: coincident detection prospects with gravitational waves}",
    eprint = "2407.04767",
    journal="",
    archivePrefix = "arXiv",
    primaryClass = "astro-ph.HE",
    month = "7",
    year = "2024"
}

@article{Mukhopadhyay:2025tvz,
    author = "Mukhopadhyay, Mainak and Kimura, Shigeo S.",
    title = "{Electromagnetic signatures from pulsar remnants of binary neutron star mergers: prospects for unique identification using multi-wavelength signatures}",
    eprint = "2506.09157",
    archivePrefix = "arXiv",
    primaryClass = "astro-ph.HE",
    month = "6",
    year = "2025"
}

@article{KM3NeT:2025npi,
    author = "Aiello, S. and others",
    collaboration = "KM3NeT",
    title = "{Observation of an ultra-high-energy cosmic neutrino with KM3NeT}",
    doi = "10.1038/s41586-024-08543-1",
    journal = "Nature",
    volume = "638",
    number = "8050",
    pages = "376--382",
    year = "2025"
}

@article{KM3NeT:2025vut,
    author = "Adriani, O. and others",
    collaboration = "KM3NeT",
    title = "{On the potential cosmogenic origin of the ultra-high-energy event KM3-230213A}",
    eprint = "2502.08508",
    archivePrefix = "arXiv",
    primaryClass = "astro-ph.HE",
    month = "2",
    year = "2025",
    journal=""
}

@article{KM3NeT:2025bxl,
    author = "Adriani, O. and others",
    collaboration = "KM3NeT, MessMapp Group, Fermi-LAT, Owens Valley Radio Observatory 40-m Telescope Group, SVOM",
    title = "{Characterising Candidate Blazar Counterparts of the Ultra-High-Energy Event KM3-230213A}",
    eprint = "2502.08484",
    archivePrefix = "arXiv",
    primaryClass = "astro-ph.HE",
    month = "2",
    year = "2025",
    journal=""
}

@article{KM3NeT:2025aps,
    author = "Adriani, O. and others",
    collaboration = "KM3NeT",
    title = "{On the Potential Galactic Origin of the Ultra-High-Energy Event KM3-230213A}",
    eprint = "2502.08387",
    archivePrefix = "arXiv",
    primaryClass = "astro-ph.HE",
    month = "2",
    journal="",
    year = "2025"
}

@article{KM3NeT:2025ccp,
    author = "Adriani, O. and others",
    collaboration = "KM3NeT",
    title = "{The ultra-high-energy event KM3-230213A within the global neutrino landscape}",
    eprint = "2502.08173",
    archivePrefix = "arXiv",
    primaryClass = "astro-ph.HE",
    month = "2",
    year = "2025",
    journal=""
}

@article{PierreAuger:2022atd,
    author = "Halim, A. Abdul and others",
    collaboration = "Pierre Auger",
    title = "{Constraining the sources of ultra-high-energy cosmic rays across and above the ankle with the spectrum and composition data measured at the Pierre Auger Observatory}",
    eprint = "2211.02857",
    archivePrefix = "arXiv",
    primaryClass = "astro-ph.HE",
    reportNumber = "FERMILAB-PUB-22-876-AD-PPD-SCD-TD",
    doi = "10.1088/1475-7516/2023/05/024",
    journal = "JCAP",
    volume = "05",
    pages = "024",
    year = "2023"
}

@article{PierreAuger:2023htc,
    author = "Halim, A. Abdul and others",
    collaboration = "Pierre Auger",
    title = "{Constraining models for the origin of ultra-high-energy cosmic rays with a novel combined analysis of arrival directions, spectrum, and composition data measured at the Pierre Auger Observatory}",
    eprint = "2305.16693",
    archivePrefix = "arXiv",
    primaryClass = "astro-ph.HE",
    reportNumber = "FERMILAB-PUB-24-0135-AD-CSAID-PPD-TD-V",
    doi = "10.1088/1475-7516/2024/01/022",
    journal = "JCAP",
    volume = "01",
    pages = "022",
    year = "2024"
}

@article{vanVelzen:2021zsm,
    author = "van Velzen, Sjoert and others",
    title = "{Establishing accretion flares from supermassive black holes as a source of high-energy neutrinos}",
    eprint = "2111.09391",
    archivePrefix = "arXiv",
    primaryClass = "astro-ph.HE",
    doi = "10.1093/mnras/stae610",
    journal = "Mon. Not. Roy. Astron. Soc.",
    volume = "529",
    number = "3",
    pages = "2559--2576",
    year = "2024"
}

@article{Mukhopadhyay:2023mld,
    author = "Mukhopadhyay, Mainak and Bhattacharya, Mukul and Murase, Kohta",
    title = "{Multimessenger signatures of delayed choked jets in tidal disruption events}",
    eprint = "2309.02275",
    archivePrefix = "arXiv",
    primaryClass = "astro-ph.HE",
    doi = "10.1093/mnras/stae2080",
    journal = "Mon. Not. Roy. Astron. Soc.",
    volume = "534",
    number = "2",
    pages = "1528--1540",
    year = "2024"
}

@article{Glaser:2019cws,
    author = "Glaser, Christian and others",
    title = "{NuRadioMC: Simulating the radio emission of neutrinos from interaction to detector}",
    eprint = "1906.01670",
    archivePrefix = "arXiv",
    primaryClass = "astro-ph.IM",
    doi = "10.1140/epjc/s10052-020-7612-8",
    journal = "Eur. Phys. J. C",
    volume = "80",
    number = "2",
    pages = "77",
    year = "2020"
}

@article{RNO-G:2020rmc,
    author = "Aguilar, J. A. and others",
    collaboration = "RNO-G",
    title = "{Design and Sensitivity of the Radio Neutrino Observatory in Greenland (RNO-G)}",
    eprint = "2010.12279",
    archivePrefix = "arXiv",
    primaryClass = "astro-ph.IM",
    doi = "10.1088/1748-0221/16/03/P03025",
    journal = "JINST",
    volume = "16",
    number = "03",
    pages = "P03025",
    year = "2021",
    note = "[Erratum: JINST 18, E03001 (2023)]"
}

@article{GRAND:2018iaj,
    author = "\'Alvarez-Mu\~niz, Jaime and others",
    collaboration = "GRAND",
    title = "{The Giant Radio Array for Neutrino Detection (GRAND): Science and Design}",
    eprint = "1810.09994",
    archivePrefix = "arXiv",
    primaryClass = "astro-ph.HE",
    doi = "10.1007/s11433-018-9385-7",
    journal = "Sci. China Phys. Mech. Astron.",
    volume = "63",
    number = "1",
    pages = "219501",
    year = "2020"
}

@article{IceCube-Gen2:2021rkf,
    author         = "S. Hallmann and Brian Clark and Christian Glaser and Daniel Smith and others",
    collaboration = "IceCube-Gen2",
    title = "{Sensitivity studies for the IceCube-Gen2 radio array}",
    eprint = "2107.08910",
    archivePrefix = "arXiv",
    primaryClass = "astro-ph.HE",
    reportNumber = "PoS-ICRC2021-1183",
    doi = "10.22323/1.395.1183",
    journal = "PoS",
    volume = "ICRC2021",
    pages = "1183",
    year = "2021"
}

@inproceedings{AlvesBatista:2024czs,
    author = "Alves Batista, Rafael",
    title = "{The Quest for the Origins of Ultra-High-Energy Cosmic Rays}",
    booktitle = "{28th European Cosmic Ray Symposium}",
    eprint = "2412.17201",
    archivePrefix = "arXiv",
    primaryClass = "astro-ph.HE",
    month = "12",
    year = "2024"
}

@article{Wittkowski:2017okb,
    author = "Wittkowski, David",
    collaboration = "Pierre Auger",
    title = "{Reconstructed properties of the sources of UHECR and their dependence on the extragalactic magnetic field}",
    doi = "10.22323/1.301.0563",
    journal = "PoS",
    volume = "ICRC2017",
    pages = "563",
    year = "2018"
}

@article{Wissel:2020sec,
    author = "Wissel, Stephanie and others",
    title = "{Prospects for high-elevation radio detection of \ensuremath{>}100 PeV tau neutrinos}",
    eprint = "2004.12718",
    archivePrefix = "arXiv",
    primaryClass = "astro-ph.IM",
    doi = "10.1088/1475-7516/2020/11/065",
    journal = "JCAP",
    volume = "11",
    pages = "065",
    year = "2020"
}

@article{Loeb:2006tw,
    author = "Loeb, Abraham and Waxman, Eli",
    title = "{The Cumulative background of high energy neutrinos from starburst galaxies}",
    eprint = "astro-ph/0601695",
    archivePrefix = "arXiv",
    doi = "10.1088/1475-7516/2006/05/003",
    journal = "JCAP",
    volume = "05",
    pages = "003",
    year = "2006"
}

@article{Lacki:2010vs,
    author = "Lacki, Brian C. and Thompson, Todd A. and Quataert, Eliot and Loeb, Abraham and Waxman, Eli",
    title = "{On The GeV \& TeV Detections of the Starburst Galaxies M82 \& NGC 253}",
    eprint = "1003.3257",
    archivePrefix = "arXiv",
    primaryClass = "astro-ph.HE",
    doi = "10.1088/0004-637X/734/2/107",
    journal = "Astrophys. J.",
    volume = "734",
    pages = "107",
    year = "2011"
}

@article{Senno:2015tra,
    author = "Senno, Nicholas and M\'esz\'aros, Peter and Murase, Kohta and Baerwald, Philipp and Rees, Martin J.",
    title = "{Extragalactic star-forming galaxies with hypernovae and supernovae as high-energy neutrino and gamma-ray sources: the case of the 10 TeV neutrino data}",
    eprint = "1501.04934",
    archivePrefix = "arXiv",
    primaryClass = "astro-ph.HE",
    doi = "10.1088/0004-637X/806/1/24",
    journal = "Astrophys. J.",
    volume = "806",
    number = "1",
    pages = "24",
    year = "2015"
}

@article{Murase:2006mm,
    author = "Murase, Kohta and Ioka, Kunihito and Nagataki, Shigehiro and Nakamura, Takashi",
    title = "{High Energy Neutrinos and Cosmic-Rays from Low-Luminosity Gamma-Ray Bursts?}",
    eprint = "astro-ph/0607104",
    archivePrefix = "arXiv",
    reportNumber = "SLAC-PUB-11954",
    doi = "10.1086/509323",
    journal = "Astrophys. J. Lett.",
    volume = "651",
    pages = "L5--L8",
    year = "2006"
}

@article{Ishihara:2019aao,
    author = "Ishihara, Aya",
    collaboration = "IceCube",
    title = "{The IceCube Upgrade - Design and Science Goals}",
    eprint = "1908.09441",
    archivePrefix = "arXiv",
    primaryClass = "astro-ph.HE",
    reportNumber = "PoS-ICRC2019-1031",
    doi = "10.22323/1.358.1031",
    journal = "PoS",
    volume = "ICRC2019",
    pages = "1031",
    year = "2021"
}

@article{Rodriguez:2024txg,
    author = "Rodr{\'\i}guez, {\'O}smar and Nakar, Ehud and Maoz, Dan",
    title = "{Stripped-envelope supernova light curves argue for central engine activity}",
    eprint = "2404.10846",
    archivePrefix = "arXiv",
    primaryClass = "astro-ph.HE",
    doi = "10.1038/s41586-024-07262-x",
    journal = "Nature",
    volume = "628",
    number = "8009",
    pages = "733--735",
    year = "2024"
}

@article{IceCubeCollaborationSS:2025jbi,
    author = "Abbasi, R. and others",
    collaboration = "(IceCube Collaboration){\textsection}, IceCube",
    title = "{Search for Extremely-High-Energy Neutrinos and First Constraints on the Ultrahigh-Energy Cosmic-Ray Proton Fraction with IceCube}",
    eprint = "2502.01963",
    archivePrefix = "arXiv",
    primaryClass = "astro-ph.HE",
    doi = "10.1103/PhysRevLett.135.031001",
    journal = "Phys. Rev. Lett.",
    volume = "135",
    number = "3",
    pages = "031001",
    year = "2025"
}

@article{AbdulHalim:2023SN,
  author = "Abdul Halim, Adila and Abreu, Pedro and Aglietta, Marco and others",
  title = "{Latest results from the searches for ultra-high-energy photons and neutrinos at the Pierre Auger Observatory}",
  doi = "10.22323/1.444.1488",
  journal = "PoS",
  year = 2023,
  volume = "ICRC2023",
  pages = "1488"
}

@article{Muzio:2019leu,
    author = "Muzio, Marco Stein and Unger, Michael and Farrar, Glennys R.",
    title = "{Progress towards characterizing ultrahigh energy cosmic ray sources}",
    eprint = "1906.06233",
    archivePrefix = "arXiv",
    primaryClass = "astro-ph.HE",
    doi = "10.1103/PhysRevD.100.103008",
    journal = "Phys. Rev. D",
    volume = "100",
    number = "10",
    pages = "103008",
    year = "2019"
}

@article{BEACON:2025qcq,
    author = "Zeolla, Andrew and others",
    collaboration = "BEACON",
    title = "{Sensitivity of BEACON to ultra-high energy diffuse and transient neutrinos}",
    eprint = "2504.13271",
    archivePrefix = "arXiv",
    primaryClass = "astro-ph.HE",
    doi = "10.1088/1475-7516/2025/09/033",
    journal = "JCAP",
    volume = "09",
    pages = "033",
    year = "2025"
}

@article{TAMBO:2025jio,
    author = {Arg{\"u}elles, Carlos A. and others},
    collaboration = "TAMBO",
    title = "{TAMBO: A Deep-Valley Neutrino Observatory}",
    eprint = "2507.08070",
    archivePrefix = "arXiv",
    primaryClass = "astro-ph.HE",
    month = "7",
    year = "2025"
}

@ARTICLE{2004AJ....127.1531H,
       author = {{Harris}, Jason and {Zaritsky}, Dennis},
        title = "{The Star Formation History of the Small Magellanic Cloud}",
      journal = {Astronomical Journal},
         year = 2004,
        month = mar,
       volume = {127},
       number = {3},
        pages = {1531-1544},
          doi = {10.1086/381953},
archivePrefix = {arXiv},
       eprint = {astro-ph/0312100},
 primaryClass = {astro-ph},
       adsurl = {https://ui.adsabs.harvard.edu/abs/2004AJ....127.1531H}
}

@article{Cuoco:2007aa,
    author = "Cuoco, A. and Hannestad, S.",
    title = "{Ultra-high energy Neutrinos from Centaurus A and the Auger hot spot}",
    eprint = "0712.1830",
    archivePrefix = "arXiv",
    primaryClass = "astro-ph",
    doi = "10.1103/PhysRevD.78.023007",
    journal = "Phys. Rev. D",
    volume = "78",
    pages = "023007",
    year = "2008"
}

@article{Mbarek:2024nvv,
    author = "Mbarek, Rostom and Caprioli, Damiano and Murase, Kohta",
    title = "{Ultrahigh-energy neutrinos as a probe of espresso-shear acceleration in jets of Centaurus A}",
    eprint = "2410.05696",
    archivePrefix = "arXiv",
    primaryClass = "astro-ph.HE",
    doi = "10.1103/PhysRevD.111.023024",
    journal = "Phys. Rev. D",
    volume = "111",
    number = "2",
    pages = "023024",
    year = "2025"
}

@article{Kachelriess:2008qx,
    author = "Kachelriess, M. and Ostapchenko, S. and Tomas, R.",
    title = "{High energy radiation from Centaurus A}",
    eprint = "0805.2608",
    archivePrefix = "arXiv",
    primaryClass = "astro-ph",
    doi = "10.1088/1367-2630/11/6/065017",
    journal = "New J. Phys.",
    volume = "11",
    pages = "065017",
    year = "2009"
}
\bibliographystyle{jhep}
\end{document}